\documentstyle[12pt,a4,epsf]{article}
\begin{document}
\setlength{\textheight}{9.4in}
\setlength{\topmargin}{-0.6in}
\setlength{\oddsidemargin}{0.10in}
\setlength{\evensidemargin}{0.4in}
\renewcommand{\thefootnote}{\fnsymbol{footnote}}

\begin{center}
{\Large {\bf D--Branes and their Absorptivity in Born--Infeld Theory}}
\vspace{0.8cm}

\vspace{0.8 cm}

{D.K. Park\raisebox{0.8ex}{\small a,b}\footnote[1]
{Email:dkpark@genphys.kyungnam.ac.kr},
S.N. Tamaryan\raisebox{0.8ex}{\small a,c}\footnote[2]{Email:
sayat@moon.yerphi.am},
H.J.W. M\"uller--Kirsten\raisebox{0.8ex}{\small a}\footnote[3]
{Email:mueller1@physik.uni-kl.de} and
 Jian-zu Zhang\raisebox{0.8ex}{\small a,d}\footnote[3]{Email:jzzhang
@online.sh.cn}

\raisebox{0.8 ex}{\small a)}{\it Department of Physics,
 University of Kaiserslautern, D--67653
Kaiserslautern, Germany}

\raisebox{0.8 ex}{\small b)}{\it Department of Physics,
Kyungnam University, Masan, 631--701, Korea}

\raisebox{0.8 ex}{\small c)}{\it Theory Department, Yerevan
Physics Institute,Yerevan 36,375036, Armenia}

\raisebox{0.8 ex}{\small d)}{\it School of Science, East China
 University of Science and Technology,
Shanghai 200237, P.R. China}}

\end{center}
\vspace{0.3cm}

{\centerline {\bf Abstract}}

Standard methods of nonlinear dynamics are used to
 investigate the stability of particles,
branes and $D$-branes of abelian Born--Infeld theory.
In particular the equation of small fluctuations
about the $D$--brane is derived and
converted into a modified Mathieu equation
and -- complementing
earlier low--energy investigations
in the case of the dilaton--axion system --
 studied in the
high--energy  domain.
 Explicit expressions are derived
for the S--matrix and absorption and reflection
amplitudes of the scalar fluctuation in the presence
of the D-brane. The results confirm physical expectations
and numerical studies of others. 
With the derivation and use of the 
(hitherto practically unknown) high energy expansion
of the Floquet exponent our considerations
also close a gap in earlier treatments
of the Mathieu equation.

\vspace{0.7cm}

\section{Introduction}

Recently Born--Infeld gauge theory has attracted considerable interest as the bosonic light--brane
approximation or limit of superstring
 theory\cite{1}, and has turned out to be a simple and 
transparent model in this context \cite{2}.
Branes, defined as extended objects in spacetime, can be fundamental or
solitonic. The connection
of these branes with a $U(1)$ gauge field was motivated by the presence of
this field in the massless part of the spectrum of open strings,
and by realising that branes with open
strings attached to them which satisfy Dirichlet boundary
conditions, or more generally one brane attached to
another, can become classically stable, solitonic objects.
It is for this reason that the dynamics of $D$-branes \cite{3} in Born--Infeld
theory is being studied in detail and
generalised \cite{4},\cite{5},\cite{6},\cite{7},\cite{8}.  Since, in general, a brane may or
may not be a solitonic configuration or BPS state, the exploration of this
question deserves particular attention. It is often stated
that a brane is BPS in view of the vanishing of a fraction of the supersymmetry
variation of the associated gaugino field.  However, since BPS states
(as classically and topologically stable states)
and Bogomol'nyi bounds have been studied in great detail in a 
host of other theories, and the approach in these is practically
standard, one would like to understand aspects of Born--Infeld
particles in a similar way, also because it is not
absolutely clear that $D$--branes are solitons
of string theory in precisely the same
way as more familiar topological solitons in field theory.
  Therefore our first intention in the following is
to study Born--Infeld particles with standard methods
of nonlinear dynamics in the simplest case of a flat
spacetime.  We begin with the free
Born--Infeld particles, i.e. BIon and catenoid \cite{4}.  Using a scale
transformation argument \cite{9} we show that these static configurations
-- which differ from ordinary solitons of nonlinear theories in
requiring a special consideration of source terms or boundary conditions 
(cf. also \cite{10},\cite{11}) --
require the number of space dimensions $p$ to be larger than 2.
We  assume spherical symmetry
and study the local stability of these configurations
by considering the second variational derivatives of
their respective actions.
Our conditions for stability are a) that the eigenfunctions
of the corresponding operator be square integrable, and  b) that
the charge $e$ be fixed,  with angular fluctuations ignored.
We then consider the case of the scalar field corresponding to
a single transverse coordinate coupled to the gauge field
(here only the electric component), i.e.
the catenoid or brane with associated open fundamental string.
We distinguish between two types of arguments in
deriving the linearised fluctuation
equation, and infer the stability of this stringy $D$-brane.
In ref.\cite{12} an explicit and detailed
 consideration of the Bogomol'nyi bound in a special
model of Born--Infeld theory has been given where the central charge
of the supersymmetry algebra plays the role of the topological
or winding number of ordinary solitons.

Our second intention in the following is the explicit
study of the small fluctuation equation about
the $D3$--brane in the high energy domain.  This
equation with singular potential 
has the remarkable property of being convertible
into a modified Mathieu equation which depends
only on one coupling parameter which is
a product of energy and electric charge.  The $S$--matrix
for scattering of the fluctuation off the brane
can be obtained in explicit form.
The $D3$--brane is therefore one of the very rare
examples allowing a detailed study of its
properties with explicit expressions for all
relevant physical quantities in both
low and high energy domains. We
 therefore expect that also $S$--duality can be uncovered
and studied in this case  (although we do not attempt
this here). Various other $Dp$--brane models have
been discovered recently whose small
fluctuation equations can be reduced to modified
Mathieu equations \cite{13,14,15} which have then
been investigated mainly by computational methods.
For the AdS/CFT correspondence the logarithmic corrections
to the low energy absorption probability are of particular interest,
since these permit a direct relation to the
discontinuity of the cut in the correlation
function of the dual two--dimensional quantum
field theory.  The first such logarithmic
correction to the absorption probability was originally
obtained in refs.\cite{kleb97,kleb98,kleb99} without
resorting to the use of Mathieu functions.
Subsequently the authors of ref. \cite{13}
considered the modified Mathieu equation and used
computational methods to generate explicit
series expansions up to several orders for the low energy absorption
probability.  In \cite{16} a different choice of expansions
was considered to obtain leading expressions more easily.
It is natural to supplement such investigations
by exploring also the high energy case, the first such consideration
being that of ref.\cite{15}.
The analytical high energy results obtained
in the following and the complementary
low energy results of ref. \cite{16}
(we also demonstrate how the
$S$--matrices are related)
are therefore directly applicable to these.
Singular potentials have been studied from time to time,
and have mostly been discarded as pathological.
It seems, however, that their real significance
lies in the context of curved spaces with
black--hole type of absorption 
\cite{singpot}.

Sections 2 and 3 deal with the BIon and the catenoid,
sections 4 and 5 with the Bogomol'nyi limit of the 
$D3$--brane and the derivation of
the linearised fluctuation equation about it.  In section 6
we consider this equation in detail in the
high energy domain and calculate
the rate of absorption of partial
waves of the fluctuation  field by the brane.
That this absorption occurs is attributed to the singularity of
the potential.
The absorptivity part of the
 paper may be looked at as the high energy complement
to the low energy case of ref.\cite{16} with the same
expression of the $S$--matrix.  All these calculations
require a matching of wave functions.  In the low
energy $S$--wave case simple considerations of
Bessel and Hankel functions suffice as was shown in refs.
\cite{kleb}.  The low energy limit is, in fact, independent
of the choice of matching point, as was shown
recently \cite{park}.  Our considerations
here, however, are general.

\vspace{1cm}

\section{The BIon}

We consider first purely static cases and write the Lagrangian
of the static BIon in $p+1$ spacetime dimensions (cf.\cite{4})
\begin{equation}
L = \int d^{p}x{\cal L}, \;\; {\cal L} = 1 -\sqrt{1-(\partial_i\phi)^2}
-\Sigma_p e \phi\delta({\bf r}),\;\;\Sigma_p=\frac{p\pi^{\frac{p}{2}}}{(\frac{p}{2})!}
\label{1}
\end{equation}
($i=1,\cdot\cdot\cdot, p$) with the charge $e$ held fixed by the constraint
\begin{equation}
e+\frac{1}{\Sigma_p}\int d\sigma_i \frac{\partial_i\phi}{\sqrt{1-(\partial_j\phi )^2}} = 0
\label{2}
\end{equation}
Eq. (\ref{1}) is the Lagrangian one obtains from the world brane action
of the pure Born--Infeld $U(1)$ electromagnetic action reduced to the purely
electric case with field $E_i=\partial_0A_i-\partial_iA_0$ and no transverse
coordinate.  The field $A_{\mu}$ is assumed to depend on the world brane
coordinates $x_{\mu}, \mu=0,\cdot\cdot\cdot,p$.
The static BIon equation of motion is 
\begin{equation}
\partial_i \left(\frac{\partial_i \phi}{\sqrt{1-(\partial_i \phi)^2}}\right) = -\Sigma_p e\delta({\bf r})
\label{3}
\end{equation}
In the
special case $p=3$ the
classical $SO(3)$ symmetric solution, called a BIon, is given by
\begin{equation}
\phi_c(r) = \int^{\infty}_r \frac{dx}{\sqrt{1 + \frac{x^4}{e^2}}} =\phi_c(0)-\int^r_0\frac{dx}{\sqrt{1+\frac
{x^4}{e^2}}}\stackrel{r\rightarrow 0}{\simeq}\left[\phi_c(0)-r+\frac{r^5}{10e^2}
\right]
\label{4}
\end{equation}
and $\phi_c(0)  = \frac{1}{4}B(\frac{1}{4},\frac{1}{4}).e^{\frac{1}{2}}
= 1.854074677.e^{\frac{1}{2}}$, $B$ being
the Bernoulli function. It is easily verified that this
solution satisfies the constraint (\ref{2}) for any value of $r$.
Defining ${\bf E} = - {\bf \nabla}\phi_c$
(so that $\phi_c = A_0$ with $\partial A_0(x_i,t)/\partial t = 0$ in the static case), and defining
${\bf D} = \frac{\partial{\cal L}}{\partial {\bf E}} =  \frac{\bf E}{\sqrt{1-{\bf E}^2}}$ we have
(with $F_{0i} =  E_i$)
\begin{equation}
T_{00}= F_{0i}\frac{\partial{\cal L}}{\partial F_{0i}} - {\cal L} 
= {\bf E}\cdot{\bf D} -{\cal L} = \frac{1}{\sqrt{1-{\bf E}^2}} - 1 + 4\pi e\phi\delta({\bf r})
\label{5}
\end{equation}
The energy $H_c$ of the BIon (obtained by integration over ${\bf R}^3$)  
is then found to be finite, i.e.
\begin{equation}
H_c = \int d{\bf x}  T_{00} =  4\pi (3.09112).e^{\frac{3}{2}}
\label{6}
\end{equation}
and in $p$ dimensions the total energy of the BIon scales correspondingly as
$e^{\frac{p}{p-1}}$.
The finiteness of the energy depends on the minus sign in (\ref{1})
and so with (\ref{3}) on the relation
\begin{equation}
\sqrt{1-\left (\phi_c^{\prime}\right)^2} = -\frac{r^2}{e}\phi_c^{\prime}
=\frac{r^2}{e\sqrt{1+\frac{r^4}{e^2}}}
\label{7}
\end{equation}
for $ 0\leq r\leq \infty$.
It may be noted that by defining ${\bf D}$ such that the left hand side of eq.(\ref{3}) is
$\partial_iD_i$, the singularity of the right hand side is associated with ${\bf D}$
rather than with ${\bf E}$ which is the decisive difference between Maxwell
and Born--Infeld electrodynamics.  A similar observation applies to the catenoid equation below.
The energy of the BIon is seen to be independent of its position
which hints at the existence of some kind of collective coordinate.
However, exploring this point further is expected to be difficult
since a moving charge generates a magnetic field,
and hence the electric field alone would not suffice.

We can use a scaling argument \cite{9} to show that here finite energy configurations require
$p$ to be larger than or equal to $3$. Under a scale transformation $x\rightarrow x^{\prime}
=\lambda x,  \phi(x)\rightarrow \phi_{\lambda}(x) = \phi(\lambda x),  \partial_i\phi(x)
\rightarrow [\partial_i\phi(x)]_{\lambda} = \lambda\partial_i\phi(\lambda x)$.
The charge $e$ defined by the constraint (\ref{2}) also changes under
the scale transformation, i.e.
\begin{equation}
e\rightarrow e_{\lambda}= - \frac{1}{\lambda^{p-2}\Sigma_p}\int\frac{d\sigma_i \partial_i\phi}
{\sqrt{1-\lambda^2(\partial_j\phi)^2}}
\label{8}
\end{equation} 
In particular for $p=3$ and radial symmetry
\begin{equation}
e^{(p=3)}_{\lambda}=\frac{r^2}{\lambda}\frac{1}{\sqrt{1-\lambda^2+\frac{r^4}{e^2}}}
\label{9}
\end{equation}
and for arbitrary values of $\lambda$ the $r$--dependence drops out only if
the limit $r\rightarrow\infty$ is taken in the evaluation of the integral. Then
\begin{equation}
\frac{e^{(p=3)}_{\lambda}}{e}\stackrel{r\rightarrow\infty}{\longrightarrow}\frac{1}{\lambda}
\label{10}
\end{equation}
But also $e_{\lambda = 1}=e$ for any $r$.
If $\phi_c$ is stable and $\neq 0$, the energy must
be stationary for $\lambda = 1$, i.e.
 $(\partial H_c/\partial\lambda)_{\lambda = 1} =0$. From this
 one finds that $p\geq 3$. Also
$(\partial^2H_c/\partial\lambda^2)_{\lambda = 1} > 0$ for $p\geq 3$.
Eqs.(\ref{6}) and (\ref{10}) show that changing the scale changes
both the charge and the energy, i.e. if the charge were variable, 
one could lower the energy and hence the configuration could be
unstable.  But fixing the charge (e.g. by a quantisation
condition) no instability is implied by the scaling condition.

We investigate the stability of the BIon further in
the special and exemplary case of $p = 3$ by
considering the second functional variation of the static
Lagrangian evaluated at $\phi_c(r)$. This can be written
and simplified in the following form (ignoring
total divergences on the way)
\begin{equation}
\delta^2 L = \frac{1}{2}\int d^3x \delta\phi {\hat A}\delta\phi,
\label{11}
\end{equation}
where
\begin{eqnarray}
{\hat A}&=&-\partial_i\frac{1}{[1-(\partial_j\phi_c)^2]^{1/2}}\partial_i
-\partial_i\frac{\partial_i\phi_c \partial_j\phi_c}{[1-(\partial_k\phi_c)^2
]^{3/2}}\partial_j\nonumber\\
&=&-\frac{1}{r^2}\frac{d}{dr}\frac{r^2}{(1-{\phi_c^{\prime}}^2)^{3/2}}
\frac{d}{dr}
\label{12}
\end{eqnarray}
The operator ${\hat A}$ can also be written
\begin{equation}
{\hat A} = -\frac{1}{(1-{\phi_c^{\prime}}^2)^{3/2}}
\left\{\frac{1}{r^2}\frac{d}{dr}r^2\frac{d}{dr}
-\frac{6}{r}{\phi_c^{\prime}}^2\frac{d}{dr}\right\}
\label{13}
\end{equation}
The classical stability of $\phi_c$ is therefore decided by
the spectrum $\{\omega_n\}$ of the small fluctuation equation
\begin{equation}
-\frac{1}{r^2}\frac{d}{dr}\frac{r^2}{(1-{\phi_c^{\prime}}^2)^{3/2}}
\frac{d}{dr}\psi_n =\omega_n\psi_n
\label{14}
\end{equation}
We explore first the existence of a zero mode 
$\psi_0$, i.e. the case $\omega = 0$.
In this case \begin{equation}
\frac{r^2}{(1-{\phi_c^{\prime}}^2)^{3/2}}
\frac{d}{dr}\psi_0 = C
\label{15}
\end{equation}
and so with $\psi_0(\infty)=0$
\begin{equation}
\psi_0=-\frac{C}{e^3}\int^{\infty}_r\frac{x^4}{(1+\frac{x^4}{e^2})^{3/2}}dx
=-C\frac{\partial}{\partial e}\int^{\infty}_r
\frac{dx}{(1+\frac{x^4}{e^2})^{1/2}}=-C\frac{\partial\phi_c}{\partial e}
\label{16}
\end{equation}
The derivative of the classical configuration $\phi_c$ with respect
to the charge $e$ indicates that a perturbation
along $\partial\phi_c/\partial e$ around $\phi_c$leaves
the static action invariant, i.e. $\phi_c(e,r)$ and 
$\phi_c(e+\delta e,r)$ have the same action
since 
$$
\frac{\partial\phi_c(e+\delta e,r)}{\partial (\delta e)}{\bigg |}_{\delta e=0}
=\frac{\partial\phi_c(e,r)}{\partial e}.
$$

We now show that the operator ${\hat A}$ does not possess negative eigenvalues,
and that therefore the BIon is a classically stable configuration. 
We let $\psi_n$ be an eigenfunction of the operator ${\hat A}$. 
Then
\begin{eqnarray}
\int d^3x\psi_n{\hat A}\psi_n
&=& -4\pi\int^{\infty}_0 dr \psi_n\frac{d}{dr}\frac{r^2}{(1-{\phi_c^{\prime}}^2
)^{3/2}}\frac{d\psi_n}{dr}\nonumber\\
&=& -4\pi\int^{\infty}_0 dr
\left\{\frac{d}{dr}\psi_n\frac{r^2}{(1-{\phi_c^{\prime}}^2)^{3/2}}
\frac{d\psi_n}{dr}-\frac{r^2}{(1-{\phi_c^{\prime}}^2)^{3/2}}
\left(\frac{d\psi_n}{dr}\right)^2\right\}\nonumber\\
&=& F+4\pi\int^{\infty}_0 dr \frac{r^2}{(1-{\phi_c^{\prime}}^2)^{3/2}}
\left(\frac{d\psi_n}{dr}\right)^2 
\label{17}
\end{eqnarray}
where $F:=F(r)|^{\infty}_0$ and
\begin{equation}
F(r) = -4\pi\psi_n\frac{r^2}{(1-{\phi_c^{\prime}}^2)^{3/2}}
\frac{d\psi_n}{dr}= -4\pi e^3\frac{\psi_n}{r^4}\left(1+\frac{r^4}{e^2}\right)
^{3/2}\frac{d\psi_n}{dr}.
\label{18}
\end{equation}
The second term on the right hand side of eq.(\ref{17})
is strictly positive.
Hence nonpositive eigenvalues imply a nonvanishing
negative value of $F$.
>From the condition $\int^{\infty}_0\psi^2_nr^2dr<\infty$,
(i.e. $\psi_n\stackrel{r\rightarrow\infty}{\simeq}1/(r^{1+\epsilon}),
\epsilon > 0$)
 it follows that
$r^2\psi_nd\psi_n/dr\rightarrow 0$ with $r\rightarrow \infty$,
 so that 
$$
F(r)\stackrel{r\rightarrow\infty}{\simeq}-4\pi r^2
\psi_n\frac{d\psi_n}{dr} \rightarrow 0
$$
and $F(\infty) = 0$.
Hence
\begin{equation}
F = - F(0) \simeq 4\pi e^3\frac{\psi_n}{r^4}
\frac{d\psi_n}{dr}\bigg|_{r\rightarrow 0}
\label{19}
\end{equation}
As $r\rightarrow 0$ eq.(\ref{14}) becomes
\begin{equation}
-\frac{1}{r^2}\frac{d}{dr}\frac{1}{r^4}\frac{d}{dr}\psi_n
 = \frac{\omega_n}{e^3}\psi_n
\label{20}
\end{equation}
In the case of the zero mode
\begin{equation}
\psi_0\simeq C_1+C_2r^5,\;\;\; r\rightarrow 0
\label{21}
\end{equation}
In this case $F=20\pi e^3 C_1C_2$. For $C_1C_2< 0 $ this is in full
compliance with (\ref{16}) and (\ref{4}) from which we obtain
$$
\psi_0\simeq -C\left(\frac{B(\frac{1}{4}, \frac{1}{4})}{8e^{1/2}}
-\frac{r^5}{5e^3}\right).
$$
For $\omega_n\neq 0$ the small--r behaviour of $\psi_n$ is
\begin{equation}
\psi_n\simeq C_n\left(1-\frac{1}{24}\frac{\omega_n}{e^3}r^8 + O(r^{16})
\right)
\label{22}
\end{equation}
so that 
$$
F=-\frac{4}{3}\pi{C_n}^2\frac{1}{r^4}r^7\bigg|_{r=0}=0
$$
Thus the conclusion is that for all eigenfunctions $\psi_n$
\begin{equation}
<\psi_n|{\hat A}|\psi_n>\;\;\; \geq 0
\label{23}
\end{equation}
This inequality excludes the possibility of the existence of
negative eigenvalues. Hence the BIon is in this sense classically
stable.

\section{The catenoid}

The Lagrangian of the static catenoid in $p+1$ spacetime dimensions
and with a source term is given by (cf.\cite{4})
\begin{equation}
L=\int d^px{\cal L},\;\;
 {\cal L} = 1 - \sqrt{1 + (\partial_i  y)^2} -\Sigma_pr_0^{p-1}y\delta({\bf r})
\label{24}
\end{equation}
where the signs have been chosen such that the energy is positive.
Here the scalar 
field $y(x_i,t)$ originates from gauge field components $A_a$ for
$a=p+1, \cdot\cdot\cdot, (d-1), d =$dimension, which represent transverse
displacements of the brane; here we consider the case of only one such
transverse coordinate, i.e. $y$,
all $d-p-1$ of which are essentially Kaluza--Klein
remnants of the $d=10$ dimensional $N=1$
electrodynamics after dimensional reduction to $p+1$ dimensions.
The Euler--Lagrange equation of the static catenoid
$y_c$ (static meaning
$\partial y(x_i,t)/\partial t = 0$) is  
given by
\begin{equation}
\partial_i \left(\frac{\partial_i y_c}{\sqrt{1 + \left(y^{\prime}_c\right)^2}}\right) = \Sigma_pr_0^{p-1}\delta ({\bf r})
\label{25}
\end{equation}
so that after integration
\begin{equation}
\frac {{\bf \nabla}y_c}{\sqrt{1+ \left(y^{\prime}_c\right)^2}} = r_0^{p-1}\frac{{\bf r}}{r^p}
\label{26}
\end{equation}
or for $r\geq r_0$
\begin{equation}
y^{\prime}_c = \stackrel{-}{(+)} \frac{{r_0}^{p-1}}{\sqrt{r^{2p-2} 
- {r_0}^{2p-2}}},\;\;\;
\sqrt{1+{y^{\prime}_c}^2}=\stackrel{-}{(+)}\frac{r^{p-1}}
{\sqrt{r^{2p-2}-r_0^{2p-2}}}
\label{27}
\end{equation}
In the case of the catenoid without source term the right hand side of
eq.(\ref{26}) can be taken to originate from a boundary condition
such as ${\bf \nabla}\cdot\left(\frac{{\bf r}}{r^p}\right) = 0$.
The domain $r \leq r_0$ is the nonsingular throat region (i.e. $y_c(r_0))$ is finite).   
One may observe that the singularity on the right hand side of eq.(\ref{25}) is associated with the
entire expression on the left whereas, like $\partial_i\phi_c$ in the BIon case, so now here
${\nabla}y_c$ is finite, i.e. the $p$-brane or single throat solution
is given by
\begin{equation}
y_c(r) = \stackrel{-}{(+)}\int^{\infty}_r dr\frac{r_0^{p-1}}{\sqrt{r^{2p-2} - r_0^{2p-2}}}
\label{28}
\end{equation}
Thus $y$ is double valued. The two possible signs can be taken to define
a brane and its antibrane.
We show at the end of this section that the solution with the minus sign
is the minimum of the
action and the solution with
the plus sign the maximum of the action.
This function is finite at $r=r_0$ and can be expressed in terms of elliptic
integrals.  For $r_0=1$ it is even simpler and has the value $y_c(1)=\stackrel{-}{(+)}\frac{1}{\sqrt{2}}{\cal K}
(\frac{1}{\sqrt{2}})$ where ${\cal K}$ is the complete elliptic integral of the first
kind.  Plotted as a function of $r$, $y_c(r)$ is
a monotonically decreasing function starting from $r_0$; pictured on a 2--dimensional
space it looks like an inverted funnel (i.e. the surface swept out by a catenary with boundaries
at the openings), thus suggesting the name catenoid.
As pointed out in ref.\cite{2},
the two possible signs of the square root allow a smooth joining of one such
funnel--shaped branch to an inverted one connected by a throat of
finite thickness, the resulting structure then representing a brane--antibrane pair.
This brane--antibrane pair is joined by the throat of finite thickness $r_0$ and
finite length. In fact, we can rewrite
eq.(\ref{28}) in terms of  
$\tilde{y}_c(x) = y_c(r_0x), x=\frac{r}{r_0}$, and for the special case of $p=3$ as
\begin{eqnarray}
\tilde{y}_c(x)&=&\stackrel{-}{(+)}\int^{\infty}_x\frac{dx}
{\sqrt{x^4-1}}=\stackrel{-}{(+)}\int^{\infty}_1\frac{dx}{\sqrt{x^4-1}}
\stackrel{+}{(-)}\int^{x}_{1}\frac{dx}{\sqrt{x^4-1}}\nonumber\\
&=&\stackrel{-}{(+)}\frac{1}{\sqrt{2}}\left[{\cal K}
\left(\frac{\sqrt{2}}{2}\right)-cn^{-1}\left(\frac{1}{x},\frac{\sqrt{2}}{2}
\right)\right]
\label{29}
\end{eqnarray}
where $x>1$ and we used formulae of ref.\cite{17}.  Inverting this expression we obtain the 
periodic function
\begin{equation}
x(y)=\left[cn\left({\cal K}\left(\frac{\sqrt{2}}{2}\right)\stackrel{+}{(-)}\sqrt{2}y, \frac{\sqrt{2}}{2}\right)\right]^{-1}
\label{30}
\end{equation}
Plotting this expression with $x$ as ordinate, one obtains the picture of a cross section
through a chain of periodically recurring funnel--shaped structures to the
one side of the throat, i.e. the series $\cup\cup\cup\cup\cdot\cdot\cdot\cdot$ representing
a series of brane--antibrane pairs along the abscissa.

Proceeding as in the above case of the static BIon
and calculating the second variational derivative we obtain
\begin{equation}
\delta^2L=\frac{1}{2}\int d^px \delta y {\hat B}\delta y
\label{31}
\end{equation}
where for $r \geq r_0$
\begin{eqnarray}
{\hat B}&=& \partial_i\frac{i}{\left[1+(\partial_i y_c)^2\right]^{1/2}}
\partial_i-\partial_i\frac{\partial_i y_c\partial_j y_c}
{\left[1+(\partial_i y_c)^2\right]^{3/2}}\partial_j\nonumber\\
&=& \frac{1}{r^2}\frac{d}{dr}\frac{r^2}{\left(1+{y_c^{\prime}}^2\right)^{3/2}}
\frac{d}{dr}=\stackrel{-}{(+)}\frac{1}{r^2}\frac{d}{dr}\frac{(r^4-r_0^4)^
{3/2}}{r^4}\frac{d}{dr}
\label{32}
\end{eqnarray}
The operator ${\hat B}$ can also be written
\begin{equation}
{\hat B} = \frac{1}{(1+{y_c^{\prime}}^2)^{3/2}}
\left\{\frac{1}{r^2}\frac{d}{dr}r^2\frac{d}{dr}
+\frac{6}{r}{y_c^{\prime}}^2\frac{d}{dr}\right\}
\label{33}
\end{equation}
Since the gauge field components $A_a, a = p+1,
 \cdot\cdot\cdot, d-1$ (of which we retain only
one), are dynamical, the
Lagrangian in the nonstatic case is
\begin{equation}
{\cal L}= 1 - \sqrt{1 - (\partial_{\mu}y)(\partial^{\mu}y)} - \Sigma_pr^{p-1}_0\delta({\bf r})
\label{34}
\end{equation}
and we can obtain the same condition of stability by considering the
dynamical fluctuation $\eta$, i.e. 
$$
y(t,{\bf x})= y_c(r)+ \eta(t,{\bf x}), \;\;\; \eta = \xi(r)e^{i{\sqrt{\omega}}t}
$$
and linearising the time--dependent Euler--Lagrange equation.
The square integrable perturbations $\xi(r)$
are the socalled ``$L^2$ deformations'' of ref.\cite{4}.
The classical stability of $y_c$ is therefore decided by
the spectrum $\{{\omega}\}$ of the small fluctuation equation
\begin{equation}
\frac{1}{r^2}\frac{d}{dr}\frac{r^2}{(1+{y_c^{\prime}}^2)^{3/2}}
\frac{d}{dr}\psi
= \frac{1}{r^2}\frac{d}{dr}\left\{\stackrel{-}{(+)}
\frac{(r^4-r_0^4)^
{3/2}}{r^4}\right\}\frac{d}{dr}\psi =\omega\psi
\label{35}
\end{equation}
We explore first the existence of a zero mode 
$\psi_0$, i.e. the case $\omega = 0$.
In this case
\begin{equation}
\frac{r^2}{(1+{y_c^{\prime}}^2)^{3/2}}\frac{d}{dr}\psi_0 = C
\label{36}
\end{equation}
and so in the case $p=3$ and $r\geq r_0$
\begin{equation}
\psi_0=C\int^{\infty}_r dx\frac{x^4}{(x^4-r_0^4)^{3/2}}=\frac{C}{2r_0}\frac
{\partial y_c}{\partial r_0}
\label{37}
\end{equation}
so that $$
\psi_0 = C\frac{\partial y_c}{\partial {r_0}^2}
$$
Here again the derivative of the classical configuration $y_c$
with respect to the parameter ${r_0}^2$
is indicative of stationarity of the action in a shift
of $r_0^2$.

We now demonstrate that the operator ${\hat B}$
with the minus sign  has  no
negative eigenvalues, and that therefore the
free catenoid is a classically stable configuration like the BIon
for fixed throat radius $r_0$. Then
\begin{eqnarray}
\int d^3x \psi{\hat B}\psi
&=&4\pi\int^{\infty}_{r_0}\ dr \psi\frac{d}{dr}\frac{r^2}{(1+{y_c^{\prime}}^2)
^{3/2}}\frac{d\psi}{dr}\nonumber\\
&=& 4\pi\frac{r^2}{(1+{y_c^{\prime}}^2)^{3/2}}\psi\frac{d\psi}{dr}\bigg|^
{\infty}_{r_0} - 4\pi\int^{\infty}_{r_0} 
dr\frac{r^2}{(1+{y_c^{\prime}}^2)^{3/2}}
\left(\frac{d\psi}{dr}\right)^2\nonumber\\
&=&\stackrel{-}{(+)}4\pi\frac{(r^4-r_0^4)^{3/2}}{r^4}
\psi\frac{d\psi}{dr}\bigg|^{\infty}_{r_0}
\stackrel{+}{(-)}4\pi\int^{\infty}_{r_0}
 dr\frac{(r^4-{r_0}^2)^{3/2}}{r^4}
\left(\frac{d\psi}{dr}\right)^2\nonumber\\
\label{38}
\end{eqnarray}
where we used eq.(\ref{27}).  The second term 
is always positive if the upper sign is
chosen.  The first term vanishes
at infinity with $\int dr r^2\psi^2 < \infty $, since
$$
-4\pi \frac{(r^4-r_0^4)^{3/2}}{r^4}\psi\frac{d\psi}{dr}
\stackrel{r\rightarrow\infty}{\simeq}-4\pi r^2\psi\frac{d\psi}{dr}
\rightarrow 0.
$$
On the other hand, in the case $r\rightarrow r_0$, we have
$$
-4\pi\frac{(r^4-r_0^4)^{3/2}}{r^4}\psi\frac{d\psi}{dr}\simeq
-32\pi\sqrt{r_0}(r-r_0)^{3/2}\psi\frac{d\psi}{dr} 
$$
As $r\rightarrow r_0$  eq.(\ref{35}) becomes
\begin{equation}
-\frac{8}{r^{3/2}_0}\frac{d}{dr}(r-r_0)^{3/2}\frac{d\psi}{dr}=\omega\psi
\label{39}
\end{equation}
In the case of the zero mode $\psi_0$
with $\omega = 0$ the considerations are analogous
to those of the BIon case
and the sum of the two terms in eq.(\ref{38}) vanishes.
In the case of $\omega\neq 0$ we therefore have
$$
\psi\simeq C\left(1 - \frac{\omega}{4}{r_0}^{3/2}\sqrt{r-r_0}\right)
$$
and
$$
\lim_{r\rightarrow r_0}\;\;(r-r_0)^{3/2}\psi\frac{d\psi}{dr}
=-\frac{\omega}{8}C^2.\lim_{r\rightarrow r_0}
{r_0}^{3/2}(r-r_0)=0
$$
This proves that for all eigenfunctions $\psi$
$$
<\psi|{\hat B}|\psi>\;\;\geq 0.
$$
Thus ${\hat B}$ has no negative eigenvalues, and the free throat is
classically stable with fixed $r_0$
for the sign chosen as in eq.(\ref{38}).
Obviously the operator ${\hat B}$
with the plus sign  has
no positive eigenvalues, which means that we have the maximum of
the action. Of course,
if we change $r_0$ (and so consider a different theory),
the expectation value of ${\hat B}$ also changes.   
One should note that the free throat we discuss here
is that
with vanishing gauge field.
The double valuedness of the solution of eq.(\ref{25}) implies
that if one solution is classically stable, the
other one is not.  Thus a multi--throat solution constructed from
these by matching both
solutions, if it exists, like the brane--antibrane solution of
ref.\cite{2}, is expected to be unstable
in view of negative as well as positive eigenvalues,
and is therefore neither a maximum nor a minimum
of the action.  In fact,
 as argued in ref.\cite{4} (after eq.(132)) equilibrium
between these should not be possible.  The reason for this
is that a symmetrical configuration, symmetrical about the plane
$x_3 = 0$ for instance, implies $\partial_3y = 0$ there.
Evaluating the stress tensor element $T_{33}$
(even for vanishing gauge field), one obtains
a negative quantity which is interpreted
as implying an attractive force between the brane and its
antibrane in this symmetrically constructed
 configuration.  This is, in fact, the
general instability of this 
configuration discussed in ref.\cite{2}.

\vspace{1cm}

\section{Coupled fields: The $D$--brane in the Bogomol'nyi limit}

In the case of coupled fields $\phi$ and $y$ (the former with source, the latter without),
the Lagrangian of the static case is (cf.\cite{4})
\begin{eqnarray}
L&=&\int d^px{\cal L}, \;{\cal L} = 1-Q-\Sigma_pe\phi\delta({\bf r}),\nonumber\\
Q&=&\bigg[1-(\partial_i\phi)^2+(\partial_iy)^2+
 (\partial_i\phi.\partial_iy)^2
-(\partial_i\phi)^2(\partial_jy)^2\bigg]^{\frac{1}{2}}
\label{40}
\end{eqnarray}
>From the first variation of $L$ we obtain the coupled equations
of the fields $\phi$ and $y$, i.e. from 
\begin{eqnarray} 
\delta L =&-& \int \delta\phi\partial_i\frac{1}{Q}
[\partial_i\phi-(\nabla\phi\cdot\nabla y)\partial_i y
+(\nabla y)^2\partial_i \phi]d^px\nonumber\\
&+&\int\delta y\partial_i\frac{1}{Q}
[\partial_i y+(\nabla\phi\cdot\nabla y)\partial_i
\phi-(\nabla\phi)^2\partial_i y]d^px\nonumber\\
&-& \Sigma_pe\int\delta\phi\delta({\bf r})d^px
\label{41}
\end{eqnarray}
(ignoring total divergences).

The source term of the electric field again suggests spherical symmetry.  In deriving the two coupled
Euler--Lagrange equations one new constant $c$ (apart from $e$) arises in the
integration of the catenoid equation, i.e.
$$
\partial_r\left(r^{p-1}\frac{\partial{\cal L}}{\partial(\partial_r y)}\right) = 0 , \;\;
r^{p-1}\frac{\partial{\cal L}}{\partial(\partial_r y)} = c
$$
We have no source term of the $y$ field because, as before, the appropriate effect
is provided by the boundary condition defining the width of
the throat.
The two equations with spherical symmetry are found to be
\begin{equation}
\frac {\phi^{\prime}}{\left[1-(\phi^{\prime})^2+(y^{\prime})^2\right]^{\frac{1}{2}}}
=-\frac{e}{r^{p-1}},\;\;
\frac{-y^{\prime}}{\left[1-(\phi^{\prime})^2+(y^{\prime})^2\right]^{\frac{1}{2}}}
=\frac{c}{r^{p-1}}
\label{42}
\end{equation}
so that
\begin{equation} 
\frac{\phi^{\prime}}{y^{\prime}}=\frac{e}{c} \equiv \frac{1}{a}
\label{43}
\end{equation}
Then
\begin{equation}
(\phi^{\prime})^2 = \frac{1}{\frac{r^{2(p-1)}}{e^2}+1-a^2},\;\;\;
(y^{\prime})^2=\frac{a^2}{\frac{r^{2(p-1)}}{e^2}+1-a^2}
\label{44}
\end{equation}
Thus the family of solutions can be parametrised in terms of the single parameter $a$
as already pointed out in ref. \cite{5}. This parameter is seen to interpolate
between the two types of static solutions.
The solution $y$ of (\ref{35}) for various values of $a^2$
is now the $p$-brane, i.e.
\begin{equation}
y(r)=\stackrel{+}{(-)}ae\int^{\infty}_rdr\frac{1}{\sqrt{r^{2(p-1)}-r^{2(p-1)}_0}}
\label{45}
\end{equation}
where $r^{2(p-1)}_0 = e^2(a^2-1)$ and for the solution to
make sense we must have $a^2\geq1$.
If $ae$ in eq.(\ref{36}) is replaced by $-ae$, the expression represents
the corresponding antibrane. Taking $e^2\rightarrow 0, a^2e^2\rightarrow$ const.
the electric field is eliminated and we regain the free catenoid solution.  In
approaching the limit
$a^2\rightarrow 1$ the width of the
 throat becomes infinitesimal with nonvanishing
electric field and the configuration can
then be considered to be a fundamental string, as argued in ref.\cite{2}.
We distinguish between three cases:
\begin{eqnarray}
|a|<1&:&\nonumber\\
&&\phi=\int^{\infty}_r\frac{dx}{\sqrt{1-a^2+x^4/e^2}},
y=a\int^{\infty}_r\frac{dx}{\sqrt{1-a^2+x^4/e^2}},\nonumber\\
|a|>1&:&\nonumber\\
&&\phi=e\int^{\infty}_r\frac{dx}{\sqrt{x^4-{r_0}^4}},
y=ae\int^{\infty}_r\frac{dx}{\sqrt{x^4-{r_0}^4}}\nonumber\\
a=\pm 1&:&
\phi=\frac{e}{r}, y = \pm \frac{e}{r}
\label{46}
\end{eqnarray}
We see that for $a^2=1$ eq. (\ref{34}) becomes the first order Bogomol'nyi equation or linearised
field equation for $y$ (as in ref.\cite{2}) 
\begin{equation}
F_{0r}\pm \frac{\partial y}{\partial r}=0
\label{47}
\end{equation}
where $F_{0r} = E_c$ is the static electric field.
This is the same equation as that obtained from the vanishing of the supersymmetry
variation of the gaugino field $\Sigma$ for half the number of 16 supersymmetries (for $d=10$
and $p=3$) 
$\epsilon_+, \epsilon_-$ of $\epsilon$ for which $\delta\Sigma = 0$, i.e.
$$
\delta_+\Sigma = 0, \;\;\; \delta_-\Sigma \neq 0
$$ 
where -- as discussed in the literature \cite{18} -- $\epsilon$
is the constant spinor of the supersymmetry variation and
$\epsilon_{\pm}$ are its chiral components.
Thus $a^2=1$ implies BPS configurations, wheras
those with $a^2\neq 1$ are non--BPS.
 Taking $a^2=0$ in eq.(\ref{36}) we
regain the BIon configuration as a local minimum of the energy whereas
for vanishing electric field one expects a local maximum, i.e. a sphaleron
configuration (as pointed out in \cite{2}).

Next we investigate the second variation of the static $L$ with
spherical symmetry. We set
\begin{equation}
\delta^2L=\frac{1}{2}\int\bigg\{\delta\phi{\hat M}\delta\phi
+\delta y{\hat N}\delta y + \delta\phi {\hat L}\delta y
+\delta y {\hat L}^{\dagger}\delta\phi\bigg\}d^px
\label{48}
\end{equation}
Again ignoring total divergences one finds
\begin{eqnarray}
{\hat M} &=& -\frac{1}{r^2}\frac{d}{dr}r^2\frac{1+{y^{\prime}}^2}{Q^3}
\frac{d}{dr},\nonumber\\
{\hat N} &=& \frac{1}{r^2}\frac{d}{dr}r^2\frac{1-{\phi^{\prime}}^2}{Q^3}
\frac{d}{dr},\nonumber\\
{\hat L} &=& \frac{1}{r^2}\frac{d}{dr}r^2\frac{\phi^{\prime}y^{\prime}}{Q^3}
\frac{d}{dr}\nonumber\\
\label{49}
\end{eqnarray}
with ${\hat L}={\hat L}^{\dagger}$.
We can now rewrite  $\delta^2L$ as
\begin{equation}
\delta^2L = \frac{1}{2}\int d^3x (\delta\phi, \delta y){\hat H}
\left(
\begin{array}{c}
\delta\phi\\
\delta y\\
\end{array}
\right)
\label{50}
\end{equation}
where
\begin{equation}
{\hat H}=
\left(\begin{array}{c}\begin{array}{cc}
M&L\\L^{\dagger}&N
\end{array}
\end{array}\right)
=\frac{1}{r^2}\frac{d}{dr}r^2 h \frac{d}{dr}
\label{51}
\end{equation}
and
\begin{equation}
h=\frac{1}{Q^3}\left(\begin{array}{c}\begin{array}{cc}
-1-{y^{\prime}}^2&y^{\prime}{\phi}^{\prime}\\
y^{\prime}{\phi}^{\prime}&1-{{\phi}^{\prime}}^2\\
\end{array}\end{array}\right),\;\;{\hat H}^{\dagger}={\hat H},
\label{52}
\end{equation}
with
\begin{equation}
h^{-1}=Q\left(\begin{array}{c}\begin{array}{cc}
-1+{{\phi}^{\prime}}^2&y^{\prime}{\phi}^{\prime}\\
y^{\prime}{\phi}^{\prime}&1+{{y}^{\prime}}^2\\
\end{array}\end{array}\right),\;\;\;{\det}\; h=\frac{1}{Q^4},
\label{53}
\end{equation}
The small fluctuation equation therefore becomes
\begin{equation}
{\hat H}\psi=\frac{1}{r^2}\frac{d}{dr}r^2h\frac{d}{dr}\psi=\omega\psi
\label{54}
\end{equation}
Again we first explore the existence of a zero mode $\psi_0$ with
\begin{equation}
r^2h\frac{d}{dr}\psi_0=\left(\begin{array}{c}\alpha\\ \beta\\
\end{array}\right)
\label{55}
\end{equation}
where $\alpha$ and $\beta$ are constants.
Setting
$$\psi_0=\left(\begin{array}{c}\phi_0\\y_0\\
\end{array}\right)
$$
and evaluating 
$\psi_0$ for the solutions of eq.(\ref{46})
we obtain with
$$
\varphi=-\int^{\infty}_r{{\phi}^{\prime}}^3(x)dx
$$
the relation
\begin{equation}
\psi_0=\frac{\phi}{e}\left\{
\left(\begin{array}{c}\alpha\\-\beta\\
\end{array}\right)
-\frac{\varphi}{e}(\alpha+a\beta)
\left(\begin{array}{c}1\\a\\
\end{array}\right)\right\}
\label{56}
\end{equation}
In the BPS limit with $y^{\prime}={\phi}^{\prime}=E_c, Q=1$, the
operator ${\hat H}$ of eq.(\ref{42}) becomes
\begin{equation}
{\hat H}=\frac{1}{r^2}\frac{d}{dr}r^2
\left(\begin{array}{c}\begin{array}{cc}
-1-E^2_c&E^2_c\\
E^2_c&1-E^2_c\\
\end{array}\end{array}\right)
\frac{d}{dr}
\label{57}
\end{equation}
Setting
$$\psi_s=\left(\begin{array}{c}\delta\phi\\ \delta y\\
\end{array}\right)=\rho(x)\left(\begin{array}{c}1\\ 1\\
\end{array}\right)
$$
for an arbitrary function $\rho(x)$
we have
\begin{equation}
{\hat H}\psi_s=\frac{1}{r^2}\frac{d}{dr}r^2
{\rho}^{\prime}\left(\begin{array}{c}\begin{array}{cc}
-1-E^2_c&E^2_c\\
E^2_c&1-E^2_c\\
\end{array}\end{array}\right)
\left(\begin{array}{c}1\\ 1\\
\end{array}\right)=\frac{1}{r^2}\frac{d}{dr}r^2
{\rho}^{\prime}\left(\begin{array}{c}-1\\ 1\\
\end{array}\right)
\label{58}
\end{equation}
Thus for arbitrary $\rho(x)$, we have $\psi_s{\hat H}\psi_s = 0$
implying $\delta^2 L=0$ or $L$
constant in a specific direction about the BPS  configuration.
This behaviour may be interpreted as
 indicative of a local symmetry, in this case
of supersymmetry, and so of the cancellation of fermionic and bosonic
contributions in the one loop approximation.
Here, of course, we have no fermionic contributions
and consequently those of the two
bosonic fields have opposite signs.

\vspace{1cm}

\section{Fluctuations about the $D$--brane}

In the following we distinguish clearly between
two different types of fluctuations.
We consider the above BPS solution for the string
attached to the 3--brane as background and consider first 
a scalar field propagating in a direction
along the string and perpendicular
to the brane and its anti--brane. The linearised equation
of small fluctuations about this background is obtained
from the second variational derivative
of the action which is the standard procedure
and we therefore consider
this first (cf. also \cite{19}).
Our treatment here is somewhat
different (see below) from that in refs.\cite{19}.
The resulting fluctuation equation has also been
given in ref.\cite{2}.  It is necessary to return
to the fully time-dependent version,
i.e.
\begin{equation}
S=\frac{1}{(2\pi)^p g_s}\int d^{p+1}x\left[1-
\sqrt{-\det(\eta_{\mu\nu}+F_{\mu\nu})}
-\Sigma_pe\phi\delta({\bf r})\right]
\label{59}
\end{equation}
where in $3+1$ dimensions $F_{\mu\nu}=F_{\mu\nu}(x_0,x_1,x_2,x_3)$.
In the electrostatic case with only one scalar field $y$
we have $ A_{\mu}=(A_0,A_1,A_2,A_3,y,0,0,0,0,0), F_{0i}=E_i$
and $F_{\mu4}=\partial_{\mu}y$ for $i=1,2,3$ and $\mu=0,1,2,3$.
Then 
\begin{equation}
\det (\eta_{\mu\nu}+F_{\mu\nu})=
\left|\begin{array}{c}\begin{array}{cc}
\begin{array}{ccc}
\begin{array}{cccc}\begin{array}{ccccc}
-1&E_1&E_2&E_3&\partial_0y\\
-E_1 & 1 & 0 & 0 & \partial_1y \\
-E_2 & 0 & 1 & 0 & \partial_2y\\
-E_3 & 0 & 0 & 1 & \partial_3y\\
-\partial_0y &-\partial_1y &-\partial_2y &-\partial_3y &1
\end{array}
\end{array}\end{array}
\end{array}
\end{array}\right|
\label{60}
\end{equation}
and so
\begin{equation}
\det(\eta_{\mu\nu}+F_{\mu\nu})
=-(1-{\bf E}^2)(1+{\bf \nabla}y^2)
-({\bf E}\cdot{\bf \nabla}y)^2+(\partial_0y)^2 
\label{61}
\end{equation}
We consider first the
Lagrangian density (remembering that the relevant fields are
$A_0, A_i$ and $y$)
\begin{equation} 
{\cal L}= 1-Q, \;\;\:\; Q=\left[(1-{\bf E}^2)(1+{\bf \nabla}y^2)
+({\bf E}\cdot{\bf \nabla} y)^2-{\dot y}^2\right]^{\frac{1}{2}}
\label{62}
\end{equation}
The equations of the static BIon and the static catenoid
discussed above follow again from the first variations
\begin{eqnarray}
\frac{\partial {\cal L}}{\partial E_i}&=&\frac{1}{Q}
\left[E_i(1+{\bf \nabla}y^2)-\partial_iy({\bf E}\cdot{\bf \nabla}y)
\right],\nonumber\\
\frac{\partial {\cal L}}{\partial \partial_iy}&=&-\frac{1}{Q}
\left[\partial_iy(1-{\bf E}^2)+E_i({\bf E}\cdot{\bf \nabla}y)
\right],\nonumber\\
\frac{\partial {\cal L}}{\partial \partial_0y}&=&-\frac{1}{Q}
\label{63}
\end{eqnarray}
In the BPS background given by
\begin{equation}
\partial_iy=E_i,\;\;\;\; {\bf E}^2=({\bf \nabla}y)^2={\bf E}\cdot{\bf \nabla}y
=\frac{e^2}{r^4}\equiv E^2_c\equiv y_c^2,\;\;\;\; Q=1
\label{64}
\end{equation}
one finds
\begin{equation}
\frac{\partial^2{\cal L}}{\partial E_i\partial E_j}=(1+E^2_c)\delta_{ij},
\frac{\partial^2{\cal L}}{\partial \partial_iy\partial \partial_jy}=
-(1-E^2_c)\delta_{ij},
\frac{\partial^2{\cal L}}{\partial E_i\partial \partial_jy}=
-E^2_c\delta_{ij},
\frac{\partial^2{\cal L}}{\partial{\dot y}^2}=1
\label{65}
\end{equation}
This enables us to write (ignoring again total divergences in
shifting derivatives)
\begin{eqnarray}
\delta^2{\cal L}=
(\delta A_0, \delta A_i, \delta y)\cdot\;\;\;\;\;\;\;\;\;\;\;\;\;
\;\;\;\;\;\;\;\;\;\;\;\;\;\;\;\;\;\;\;\;\;\;
\;\;\;\;\;\;\;\;\;\;\;\;\;\;\;\;\;\;\;\;\;\;\;\;
\;\;\;\;\;\;\;\;\;\;\;\;\;\;\;\;\;\;\;\;\nonumber\\
\cdot\left(\begin{array}{c}\begin{array}{cc}\begin{array}{ccc}
-\partial_i(1+E^2_c)\partial_i&\partial_i(1+E^2_c)\partial_0&
-\partial_iE^2_c\partial_i\\
\partial_0(1+E^2_c)\partial_i&-\partial_0(1+E^2_c)\partial_0
&\partial_0E^2_c\partial_i\\
-\partial_iE^2_c\partial_i&+\partial_iE^2_c\partial_0&
\partial_i(1-E^2_c)\partial_i-\partial_0\partial_0
\end{array}
\end{array}
\end{array}\right)
\left(\begin{array}{c}
\delta A_0\\ \delta A_i\\ \delta y
\end{array}\right)
\label{66}
\end{eqnarray}
In the linear approximation the Euler--Lagrange
equations of the fluctuations
 $\delta y\equiv \eta,
 \delta E_i=\partial_0\delta A_i
-\partial_i\delta A_0$
are therefore given by the following set of three equations
\begin{eqnarray}
-\frac{d^2}{dt^2}\eta +\partial_i(1-E^2_c)\partial_i\eta
+\partial_iE^2_c(\partial_0 \delta A_i- \partial_i \delta A_0)&=&0,\\
\frac{d}{dt}(1+E^2_c)(\partial_0
 \delta A_i- \partial_i\delta A_0)
-\frac{d}{dt}E^2_c \partial_i\eta &=&0,\\
 \partial_i(1+E^2_c)(\partial_0 \delta A_i- \partial_i\delta A_0)
-\partial_iE^2_c \partial_i\eta &=& 0 
\label{67,68,69}
\end{eqnarray}
The last of these three equations can be seen to be a constraint
by appying $\partial/\partial t$ and using the second equation.
Substituting from the last
$$
\partial_iE^2_c(\partial_0 \delta A_i-\partial_i \delta A_0) = 
\partial_iE^2_c\partial_i\eta
-\partial_i(\partial_0 \delta A_i-\partial_i \delta A_0)
$$
into the first equation we obtain
\begin{equation}
-\frac{d^2}{dt^2}\eta + \triangle\eta-\partial_i(\partial_0
\delta A_i-\partial_i \delta A_0) =0
\label{70}
\end{equation}
The second of the three equations can be written in the form
\begin{equation}
(1+E^2_c)(\partial_0\delta A_i-\partial_i\delta A_0)-E^2_c\partial_i\eta=
(1+E^2_c)C_i(r)
\label{71}
\end{equation}
where ${\bf C}(r)$ is an arbitrary function.  Dividing eq.(\ref{60})
by $(1+E^2_c)$ and taking the derivative $\partial_i$, we obtain
\begin{eqnarray}
\partial_i(\partial_0\delta A_i-\partial_i\delta A_0)
&=&\partial_i\frac{E^2_c}{1+E^2_c}
\partial_i\eta +\partial_iC_i\nonumber\\
&=&\frac{E^2}{1+E^2_c}\triangle\eta+\frac{2E_cE^{\prime}_c}{(1+E^2_c)^2}
\frac{x_i}{r}\partial_i\eta + \partial_iC_i
\label{72}
\end{eqnarray}
Replacing on the right hand side $E^2_c\partial_i\eta$ by the
expression in eq.(\ref{60})this becomes
\begin{equation}
\partial_i(\partial_0\delta A_i-\partial_i\delta A_0)
=\frac{E^2_c}{1+E^2_c}\triangle\eta
+\frac{2E^{\prime}_c}
{E_c(1+E^2_c)}\frac{x_i}{r}
\left[(\partial_0\delta A_i-\partial_i\delta A_0)-C_i\right]
+\partial_iC_i
\label{73}
\end{equation}
Choosing as gauge fixing condition the relation
$$
\frac{2E^{\prime}_c}
{E_c(1+E^2_c)}\frac{x_i}{r}
\left[(\partial_0\delta A_i-\partial_i\delta A_0)-C_i\right]
+\partial_iC_i=0
$$ 
one obtains the following fluctuation equation for $\eta$
\begin{equation}
-(1+E^2_c)\frac{d^2\eta}{dt^2} + \triangle\eta = 0
\label{74}
\end{equation}
All the relations from (\ref{60}) to (\ref{74}) describe
perturbations along the string and perpendicular to
the brane. Eq. (\ref{74}) cannot be considered
independently of the others as is apparent from the linkage
of the fields in the above equations.  Thus if one wants
to determine the radiation of the string between
the brane and the antibrane, one must
connect the asymptotic behaviour of the field $\eta$ with
that of the vector field $\delta A_{\mu}$.

However, an equation like (\ref{74})is also obtained if one  
evaluates the determinant in the Born--Infeld Lagrangian
at the BPS background and with an additional time--dependent
scalar $\eta$, representing the fluctuation field
along a new spatial direction (cf. also ref.\cite{2}).
In this case this new scalar field in the $D$--brane
background has no relevance to the string
radiation, and we have
\begin{equation}
{\det (\eta_{\mu\nu}+F_{\mu\nu})|}_{BPS,\eta}=
\left|\begin{array}{c}\begin{array}{cc}
\begin{array}{ccc}
\begin{array}{cccc}\begin{array}{ccccc}\begin{array}{cccccc}
-1&E_1&E_2&E_3&0&\partial_0\eta\\
-E_1 & 1 & 0 & 0 & E_1&\partial_1\eta \\
-E_2 & 0 & 1 & 0 & E_2 &\partial_2\eta\\
-E_3 & 0 & 0 & 1 & E_3& \partial_3y\\
0 & -E_1 & -E_2 & -E_3 & 1 & 0\\
-\partial_0\eta &-\partial_1\eta &-\partial_2\eta &-\partial_3\eta & 0 & 1
\end{array}
\end{array}\end{array}
\end{array}\end{array}
\end{array}\right|
\label{75}
\end{equation}
and so
\begin{equation}
{\det(\eta_{\mu\nu}+F_{\mu\nu})|}_{BPS,\eta}
=-(1+{{\bf E}_c}^2)(\partial_0\eta)^2-(\partial_i\eta)^2-1 
\label{76}
\end{equation}
Thus the Lagrangian density becomes
\begin{equation}
{\cal L}= 1 - \sqrt{1+({\bf \nabla}\eta)^2-(1+{{\bf E}_c}^2)(\partial_0\eta)^2}
\label{77}
\end{equation}
By expanding the square root and
retaining only the lowest order terms,
 we again obtain a fluctuation equation like (\ref{65}),
but this time for $\eta$ with no relevance to
radiation of the string.
This is equivalent to studying the scattering of
the scalar $\eta$ off a corresponding supergravity
background.

\section{Absorption of scalar in background of $D3$ brane}

We now consider the equation of small fluctuations, i.e. eq.(\ref{74}),
in more detail. The fluctuation $\eta(t,{\bf x})$ represents
 a scalar field that
impinges on the brane which reflects part of it and absorbs part of it
depending on the energy $\omega$ of the field. 
The absorption results from and takes place into the singularity
of the real potential  which corresponds to the
black hole with zero event horizon in 
the analogous case of the dilaton--axion
system of e.g. ref.\cite{16}. This absorption
is a classical phenomenon.
We therefore consider the equation
\begin{equation}
\triangle_r\xi + \omega^2\left[1+\frac{e^2}{r^4}\right]\xi = 0
\label{78}
\end{equation}
One can argue that the absorption is a consequence
of the nonhermiticity of the potential.

The radial part of this equation is with $\xi=r^{-1}\Psi Y_{lm} $ and
angular momentum $l$
\begin{equation}
\frac{d^2\Psi}{dr^2} +\left[-\frac{l(l+1)}{r^2}
+\omega^2\left(1+\frac{e^2}{r^4}\right)\right]\Psi=0
\label{79}
\end{equation}
This equation is a radial Schr\"odinger equation
for an attractive singular potential $\propto
r^{-4}$ but depends only on the single coupling
parameter $\kappa = e\omega^2$ for
constant positive Schr\"odinger energy,
i.e. for 
 $S$-waves
the equation is with $x=\omega r$ simply
\begin{equation}
\left(\frac{d^2}{dx^2}+1+\frac{\kappa^2}{x^4}\right)\Psi = 0
\label{80}
\end{equation}
In the following we consider the general
case, i.e. $l\neq 0$. The simplified case
of the singular potential replaced by an effective
delta--function potential has been
considered in refs.\cite{2} and \cite{19}.
The solutions and properties of such
 equations have been studied in detail in the literature,
in both the small-- and large--$\kappa $ domains and with inclusion of the
centrifugal term $-l(l+1)/r^2$ in eq.(\ref{79}) for the calculation of Regge
trajectories $l\rightarrow \alpha_n(\omega^2)$ \cite{20},
\cite{21},\cite{22},\cite{23}.
A recent investigation which
 attempts to treat arbitrary power singular potentials is ref.\cite{24}.
Eq.(\ref{79}) describes
waves above the singular potential well.
 With the substitutions
\begin{equation}
\Psi(r)=r^{\frac{1}{2}}\psi(r),\;\; r = \sqrt{e}e^z,\;\;
 h^2=e\omega^2,\;\;  a =l+\frac{1}{2},
\label{81}
\end{equation}
the equation becomes the modified Mathieu equation
\begin{equation}
\frac{d^2\psi}{dz^2}+\left[2h^2\cosh 2z -a^2\right]\psi=0
\label{82}
\end{equation}
which
has been studied in detail in the literature \cite{25}
(though some properties, such as large--h asymptotic
expansions of Fourier coefficients, have even now
not yet been published).
Here we study the S--matrix in the domain of finite values
of angular momentum $l$ and $h^2\neq 0$, i.e. in the domain
of $h^2$ large. Relevant solutions and matching conditions
for this case have been developed in \cite{26}and \cite{27}.
We follow the latter of these references here since this
makes full use of the symmetries of the solutions.
Moreover we can determine also the Floquet exponent $\nu$
which ref.\cite{26} leaves undetermined and only remarks
that the notion that this is a known function of
(our) $a^2$ and $h^2$ is ``partly a convenient fiction''.

For convenience we set in eq. (\ref{82}) as in ref.\cite{27,28}
\begin{equation}
a^2=-2h^2+2hq+\frac{\triangle(q,h)}{8}
\label{83}
\end{equation}
where $q$ is a parameter to be determined as the solution of
this equation and $\triangle/8$ is the remainder of the
large--h asymptotic expansion (\ref{83}), the
various terms of which are determined
concurrently with corresponding iteration contributions of the  
solutions $\psi$ of the equation and are known explicitly
to many orders \cite{28}.
Then setting in eq. (\ref{82})
\begin{equation}
\psi(q,h;z) = A(q,h;z)exp[\pm 2hi\sinh z]
\label{84}
\end{equation}
we obtain an equation for $A$ which can be written
\begin{equation}
\cosh z\frac{dA}{dz} + \frac{1}{2}(\sinh z\pm iq)A=\pm\frac{1}{4hi}
\left[\frac{\triangle}{8}A-\frac{d^2A}{dz^2}\right]
\label{85}
\end{equation}
We let $A_q(z)$ be the solution of this equation when the right
hand side is replaced by zero (i.e. in the limit $h\rightarrow\infty$).
Then one finds easily
\begin{equation}
A_q(z)=\frac{1}{\sqrt{\cosh z}}
\left(\frac{1+i\sinh z}{1-i\sinh z}\right)^{\mp q/4}
\stackrel{z\rightarrow\infty}{\sim}\sqrt{2} e^{-z/2}e^{\mp i\pi q/4}
\label{86}
\end{equation}
Correspondingly the various solutions $\psi$ are
\begin{eqnarray}
\psi(q,h;z)&=&A_q(z)exp[\pm 2hi\sinh z]\stackrel{z\rightarrow\infty}{\simeq}
\frac{exp(\pm ihe^z)}{\sqrt{\cosh z}}e^{\mp i\pi q/4},\nonumber\\
\psi(q,h;z)&=&A_q(z)exp[\pm 2hi\sinh z]\stackrel{z\rightarrow-\infty}{\simeq}
\frac{exp(\mp ihe^{|z|})}{\sqrt{\cosh z}}e^{\mp i\pi q/4}
\label{87}
\end{eqnarray}
We make the important observation that given one solution
$\psi(q,h;z)$ we can obtain the linearly independent one
either as $\psi(-q,-h;z)$ or as $\psi(q,h;-z)$,  
the expression (\ref{83}) remaining unchanged.
With the solutions as they stand, of course
$\psi(q,h;z)=\psi(-q,-h;-z)$.
Below we require solutions $He^{(i)}(z), i=1,2,3,4$,
 with some specific asymptotic 
behaviour. We define these in terms of the function
\begin{equation}
Ke(q,h;z):=\frac{exp[i\pi q/4]}{\sqrt{-2ih}}A_q(z)exp[2hi\sinh z]
\equiv k(q,h)\psi(q,h;z)
\label{88}
\end{equation}
Since this function differs from a solution $\psi$ by a factor
$k(q,h)$, it is still a solution but not with
the symmetry property $\psi(q,h;z)=\psi(-q,-h;-z)$. Instead,
after performing this cycle of replacements the function
picks up a factor,
i.e.
\begin{equation}
Ke(q,h;z)=\frac{k(q,h)}{k(-q,-h)}Ke(-q,-h;-z),\;\;
\frac{k(q,h)}{k(-q,-h)}=e^{i\frac{\pi}{2}(q+1)}
\label{89}
\end{equation}
in leading order.
One can easily show that the quantity $\Phi_0$ of ref. \cite{26} is related
to $q$ by $\Phi_0=iq\pi/2 + O(1/h)$.
In Fig. 1 we show the behaviour of $q$ as a function of $h$.
 In order to be able to obtain the $S$--matrix,
we have to match a solution valid at $z=-\infty$ to a combination
of solutions valid at $z=\infty$.  This is achieved with the help
of Floquet solutions $Me_{\pm\nu}(z,h^2)$. As such,
these satisfy the same circuit relation as a solution
 $M^{(1)}_{\pm\nu}(z,h^2)$ of eq.(\ref{82}) expanded
in a series of Bessel functions, i.e. we have the
proportionality
\begin{equation} 
Me_{\nu}(z,h^2)={\alpha}_{\nu}M^{(1)}_{\nu}(z,h^2),\;\;
{\alpha}_{\nu}(h^2)=Me_{\nu}(0,h^2)/M^{(1)}_{\nu}(0,h^2)
\label{90}
\end{equation}
The functions $Me_{\pm\nu}(z,h^2)$ are expansions of the
modified (hence `M' instead of `m') Mathieu equation in terms
of exponentials (hence `e') which are uniformly
convergent in any finite domain of $z$.
For large values of the argument $2h\cosh z$
of the Bessel functions of 
the modified Mathieu function $M^{(1)}_{\nu}(z,h^2)$ 
can be reexpressed in terms of Hankel functions. With the
dominant terms of these we can obtain the large  $2h\cosh z$
asymptotic behaviour of the Floquet function $Me_{\pm\nu}(z,h^2)$,
i.e. for $|z|\rightarrow\infty$
\begin{equation}
 Me_{\pm\nu}(z,h^2)\simeq
 exp[\pm i\pi\gamma/2]\frac{\cos(2h\cosh z\mp\nu\pi/2-\pi/4)}
{\sqrt{2h\cosh z}}
\label{91}
\end{equation}
where (with $ Me_{\nu}(-z,h^2)=Me_{-\nu}(z,h^2)$)
\begin{equation}
exp[i\pi\gamma]=\frac{\alpha_{\nu}(h^2)}{\alpha_{-\nu}(h^2)}
= M^{(1)}_{-\nu}(0,h^2)/M^{(1)}_{\nu}(0,h^2)
\label{92}
\end{equation}
We now define the following set of solutions of eq.(\ref{82})
by setting
\begin{eqnarray}
He^{(2)}(z,q,h)&=&Ke(q,h,z), He^{(1)}(z,q,h)= He^{(2)}(z,-q,-h),\nonumber\\
He^{(3)}(z,q,h)&=&He^{(1)}(-z,q,h), He^{(4)}(z,q,h) = He^{(2)}(-z,q,h)
\label{93}
\end{eqnarray}
The solutions so defined have the following asymptotic behaviour
(where $\epsilon(z)=(2h\cosh z)^{-1/2}$):
\begin{eqnarray}
He^{(1)}(z,q,h)&=&\epsilon(z)\cdot exp[-ihe^z -i\frac{\pi}{4}],\;\;\;
\Re z>>0, \nonumber\\
&\stackrel{r\rightarrow \infty}{\sim}&
\frac{exp[-i\omega r -i\pi/4]}{\sqrt{\omega r}},\nonumber\\
He^{(2)}(z,q,h)&=&\epsilon(z)\cdot exp[ihe^z +i\frac{\pi}{4}],
\;\;\; \Re z>>0, \nonumber\\
&\stackrel{r\rightarrow \infty}{\sim}&
\frac{exp[i\omega r+ i\pi/4]}{\sqrt{\omega r}},
\nonumber\\
He^{(3)}(z,q,h)&=&\epsilon(z)\cdot exp[-ihe^|z| -i\frac{\pi}{4}],
\;\;\; \Re z <<0, \nonumber\\
He^{(4)}(z,q,h)&=&\epsilon(z)\cdot exp[ihe^|z| +i\frac{\pi}{4}],
\;\;\; \Re z <<0,\nonumber\\
&\stackrel{r\rightarrow 0}{\sim}&\frac{r^{1/2}exp[ie\omega/r
 +i\frac{\pi}{4}]}{(e\omega)^{1/2}},
\label{94}
\end{eqnarray}
For the following reasons we choose the latter,
 i.e. the solution $He^{(4)}(z,q,h)$, as our
solution at $r=0$. The time--dependent wave function
with this asymptotic behaviour is proportional to
$$
e^{-i\omega t +ie\omega/r+i\pi/4}
$$
Fixing the wave front by setting 
$
\varphi = -\omega t + e\omega/r +\pi/4 = const.
$  
and considering the propagation of this wave front, we have
$$
r=\frac{e\omega}{\varphi + \omega t -\pi/4}
$$
so that when $t\rightarrow\infty: r\rightarrow 0$. This means
that the origin of coordinates acts as a sink.

With eq. (\ref{91}) we therefore equate in the domain $\Re z>>0$:
\begin{eqnarray}
Me_{\nu}(z,h^2)&=&\frac{i}{2}exp[i\pi\gamma/2]\bigg\{exp[i\nu\frac{\pi}{2}]
He^{(1)}(z,q,h)-exp[-i\nu\frac{\pi}{2}]He^{(2)}(z,q,h)\bigg\},\nonumber\\
Me_{-\nu}(z,h^2)&=&\frac{i}{2}exp[-i\pi\gamma/2]\bigg\{exp[-i\nu\frac{\pi}{2}]
He^{(1)}(z,q,h)\nonumber\\
&-&exp[i\nu\frac{\pi}{2}]He^{(2)}(z,q,h)\bigg\},
\label{95}
\end{eqnarray}
where the second relation was obtained by changing the sign on $\nu$ in the
first. Changing the sign of $z$ we obtain in the domain $\Re z<<0$: 
\begin{eqnarray}
Me_{\nu}(-z,h^2)&=&Me_{-\nu}(z,h^2)\nonumber\\
&=&\frac{i}{2}exp[i\pi\gamma/2]
\bigg\{exp[i\nu\frac{\pi}{2}]
He^{(3)}(z,q,h)-exp[-i\nu\frac{\pi}{2}]He^{(4)}(z,q,h)\bigg\},\nonumber\\
Me_{\nu}(z,h^2)&=&\frac{i}{2}exp[-i\pi\gamma/2]\bigg\{exp[-i\nu\frac{\pi}{2}]
He^{(3)}(z,q,h)\nonumber\\
&-&exp[i\nu\frac{\pi}{2}]He^{(4)}(z,q,h)\bigg\},
\label{96}
\end{eqnarray}
These relations are now valid over the entire range
of $z$.
Substituting eqs.(\ref{96})into eqs.(\ref{95}) and eliminating $He^{(3)}$
we obtain
\begin{equation}
-\sin\pi\nu.He^{(4)}(z,q,h)=\sin\pi(\gamma+\nu).He^{(1)}(z,q,h)
-\sin\pi\gamma.He^{(2)}(z,q,h)
\label{97}
\end{equation}
In a similar way one obtains the relations
\begin{eqnarray}
-\sin\pi\nu.He^{(2)}(z,q,h)&=&\sin\pi(\gamma+\nu).He^{(3)}(z,q,h)
-\sin\pi\gamma.He^{(4)}(z,q,h)\nonumber\\
\sin\pi\nu.He^{(1)}(z,q,h)&=&-\sin\pi\gamma.He^{(3)}(z,q,h)
+\sin\pi(\gamma-\nu).He^{(4)}(z,q,h)\nonumber\\
\label{98}
\end{eqnarray}
>From eqs.(\ref{89}) and (\ref{93}) we see that
$He^{(2)}(z,q,h)$ is proportional to
$He^{(3)}(z,q,h)$.
>From (\ref{89}) and (\ref{98}) we see that the proportionality
factor is given by
\begin{equation}
exp[i\frac{\pi}{2}(q+1)]=-\frac{\sin\pi(\gamma+\nu)}{\sin\pi\nu}
\label{99}
\end{equation}
>From eq.(\ref{97}) we can now deduce the S--matrix $S_l\equiv e^{2i{\delta}_l}$,
where ${\delta}_l$ is the phase shift.  The latter is defined by the
following large $r$ behaviour of the solution chosen at $r=0$, which
in our case is the solution $He^{(4)}$. Thus here the S--matrix
is defined by (using (\ref{97}))
\begin{eqnarray}
&-&\sin\pi\nu\frac{r^{1/2}e^{ie\omega/r+i\pi/4}}{(e\omega)^{1/2}}\nonumber\\
&\stackrel{r\rightarrow\infty}{=}&
-(-1)^l\frac{\sin\pi(\gamma+\nu)e^{-i\pi/4}}
{\sqrt{\omega r}}\bigg[\frac{\sin\pi\gamma (-1)^l}{\sin\pi(\gamma+\nu)}
e^{i\pi/2}e^{i\omega r}-(-1)^le^{-i\omega r}\bigg]\nonumber\\
&\equiv &\frac{e^{-i\delta}e^{-il\pi/2}}{2i\sqrt r}
\bigg[S_le^{i\omega r}-(-1)^le^{-i\omega r}\bigg]
\label{100}
\end{eqnarray}
>From this we deduce that
\begin{equation}
S_l=\frac{\sin\pi\gamma}{\sin\pi(\gamma+\nu)}e^{i\pi(l+1/2)}
=-\frac{\sin\pi\gamma}{\sin\pi\nu}e^{i\pi(l-\frac{1}{2}q)}
\label{101}
\end{equation}
We can see the relation of this high--energy (i.e.
large $|h|$) expression of the S--matrix to the low--energy
expression of ref.\cite{25} by recalling that 
$R$ of the latter is here $exp(i\pi\gamma)$. With this
identification we can write $S_l$
$$
S_l=\frac{R-\frac{1}{R}}{(Re^{i\pi\nu}-\frac{e^{-i\pi\nu}}{R})}e^{i\pi(l+1/2)},
\;\;\;R\equiv e^{i\pi\gamma},
$$
which agrees with the $S$--matrix of ref.\cite{16},
i.e. we thus obtained the same exact expression of the
$S$--matrix here with our large--$h$
considerations.
In fact, comparison with the considerations given there allows one
to write down the reflection and transmission amplitudes $A_r$ and
$A_t$ as $A_r=2i\sin\pi\gamma$ and $A_t=\sin\pi\nu$ respectively.
We thus have one and the same expression  for the $S$--matrix for the
two asymptotic regions, i.e. in the low energy
and high energy domains.
One should therefore be able to proceed directly to the large--$h$ case
from the exact $S$--matrix derived in the small--$h$
domain.  This is an interesting calculation
which we do not attempt to go into here.  We only indicate
in Appendix A the first necessary step in that direction, i.e.
the derivation of large--$h$ asymptotic expansions for the
Fourier coefficients of Mathieu functions.
In this connection we make the following two observations.
1) Eq.(\ref{80}) is invariant under interchanges
$x\leftrightarrow \kappa/x, \Psi\leftrightarrow\Psi x$
which means that the inner or string region
is equivalent or dual to the outer or brane region.
2) Due to the $SL(2,R)$ invariance of the $D3$--brane
its action is mapped into that of an equivalent $D3$--brane
by $S$--duality transformations\cite{29} or, in other
words, weak--strong duality takes the $D3$--brane
into itself\cite{3}.  It would be
interesting to find some connection between these
properties, or equivalently the symmetry which the $SL(2,R)$
invariance of the $D3$--brane action imposes on the $S$--matrix.

The quantity $\gamma$ is now to be determined from eq. (\ref{99}).
One finds
\begin{equation}
\sin\pi\gamma=\sin\pi\nu\bigg\{-ie^{i\frac{\pi}{2}q}\cos\pi\nu
\pm\sqrt{1+e^{i\pi q}\sin^2\pi\nu}\bigg\}
\label{102}
\end{equation}
It remains to determine the Floquet exponent $\nu$ in terms
of $q$ and $h$.  In Appendix B we derive the appropriate 
large--$h$ behaviour of $\nu$ for the case of the periodic
Mathieu equation. Replacing there the eigenvalue $\lambda$
by
$ a=(l+\frac{1}{2})^2$ and observing that $h^2$
remains  $h^2$, the
appropriate relation for our considerations is
\begin{eqnarray}
\cos\pi\nu + 1 &=&
\frac{\pi e^{4h}}{(8h)^{q/2} }
\bigg[\frac{1+\frac{3(q^2+1)}{64h}}
{\Gamma[\frac{3}{4}-\frac{q}{4}]\Gamma[\frac{1}{4}-\frac{q}{4}]}
+O(\frac{1}{h^2})\bigg]\nonumber\\
&=& \frac{e^{4h}}{(8h)^{q/2}}\bigg[\frac{\left(1
+\frac{3(q^2+1)}{64h}\right)\Gamma(\frac{q+1}{2})
\cos(\frac{q\pi}{2})} 
{\sqrt{2\pi}2^{q/2}}+O(\frac{1}{h})\bigg]
\label{103}
\end{eqnarray}
Since the right hand side grows exponentially with increasing $h$
the Floquet exponent $\nu$ must have a large
imaginary part.  Since the right
hand side is real, the real
part of $\nu$ must be an integer.
Using Stirling's formula we can approximate the equation
for $q\simeq h$ (i.e. irrespective of what the value of
$l$ is) as
\begin{equation}
\cos\pi\nu + 1=\sqrt{\frac{h}{2}}\cos(\frac{h\pi}{2})(e^7/32)^{h/2}
\simeq\sqrt{\frac{h}{2}}e^{1.8h}\cos(\frac{h\pi}{2})
\label{104}
\end{equation}
>From eq.(\ref{101}) and eq.(\ref{102}) we obtain
\begin{equation}
S_l=ie^{il\pi}\bigg(\cos\pi\nu - \sqrt{\cos^2\pi\nu-1-e^{iq\pi}}\bigg)
\label{105}
\end{equation}
>From this we obtain the absorptivity $A(l,h)$ of the l--th
partial wave, i.e.
\begin{equation}
A(l,h):=1-|S_l|^2
\label{106}
\end{equation}
with near asymptotic behaviour
\begin{equation}
A(l,h)\simeq 1-\frac{2\pi (16h)^q}{e^{8h}\left\{\Gamma(\frac{q+1}{2})\right\}^2}
\label{107}
\end{equation}
In Figs. 2, 3 and 4 we plot $A(l,h)$ as a function of $h$. One can clearly see
the expected asymptotic approach to unity and in Fig. 2
some sign of
rapidly damped oscillations.  This behaviour agrees with
that obtained on general grounds in ref.\cite{15}.
We also observe that in the high energy limit logarithmic contributions
as in the low energy expansions, discovered
originally in \cite{kleb97,kleb98,kleb99}, and typical of the
low energy expansions of \cite{13} and \cite{16},
do not arise.
Of course, these plots do not extend down to $h=0$, since
our asymptotic solutions become meaningless in that domain.
The continuation to $h=0$ can be obtained, however, from
small--h expansions
such as those  derived in refs.\cite{13} and \cite{16}. Thus the
absorptivity $A(l,h)$ is known over the entire range of $h$.
We observe that $S_l=0$ for $q=1,3,5,\cdot\cdot\cdot$,
with $[(l+1/2)^2+2h^2]/2h\simeq 1,3,5,\cdot\cdot\cdot$. Only
in the plot for $l=2$ is $h$ sufficiently large to hint at
these zeros.

\vspace{1cm}
\section{Concluding remarks}
Branes, whether fundamental or solitonic, play an important role
in all aspects of string theory.  In particular $D$--branes
have been looked at as string--theory analogues of solitons
of simple field theories, and some of their important
properties such as charges are well understood.  Our first
objective in the above was to investigate properties of solitonic 
objects of Born--Infeld theory
 in ways familiar from field theory, in particular
their classical stability.  It was shown that the BIon and the
catenoid as distinct, i.e. free objects,
 are stable configurations
whereas the brane--antibrane system is unstable; we
also recognised the zero modes associated
with these and their significance.  We then
considered the $D3$--brane of Born--Infeld theory and recognised
this as a BPS state that preserves half of the
number of supersymmetries as discussed in detail already in \cite{2}.
  The equation of small
fluctuations about this $D3$--brane was derived and shown to be convertible
into a modified Mathieu equation.
  The low energy solutions of this equation,
 the $S$--matrix for scattering of a massless scalar off the
brane and the corresponding absorption
and reflection amplitudes are similar to
those for the dilaton--axion system investigated
first in refs.\cite{kleb97,kleb98,kleb99}, where
the important logarithmic contributions were discovered,
and then investigated 
in extensive detail in \cite{13} and \cite{16}.
Here we performed the high energy calculations which
complement in particular those of
\cite{16}, thus completing
the investigation of the modified Mathieu
equation for the purpose of obtaining absorption
cross sections for all such cases.  In particular
the behaviour of the important Floquet
exponent involved in these calculations
(in general a complex quantity) is now fully understood,
the Floquet exponent being
 vital in the evaluation of the $S$--matrix which we
derive and the calculation of the corresponding
absorption amplitudes and cross sections.
According to our findings the high energy limit of the
absorption cross section does not involve
logarithmic contributions, quite contrary
to the low energy limit.

The high energy case considered here
 is not only of interest in the immediate
context of the Born--Infeld model considered here, but together
with the low--energy case also of considerable interest in connection
with the concept of duality which links weak coupling with
strong coupling.  The $D3$--brane with Schr\"odinger
potential coupling $e\omega^2$, which links the gauge field
charge $e$ with energy$\omega$ of the incoming scalar field
is presumably the ideal example for the investigation of this
property. Investigations elucidating this aspect are
of considerable interest. We also envisage interest in the
study of non--BPS configurations, including sphalerons
and bounces, as a matter of principle, i.e. even if the effect of
these is not of dominant importance.
 Finally we remark that it should be possible
to proceed directly from the $S$--matrix derived in ref.\cite{16}
to the high--energy case here by using appropriate asymptotic
expansions for the cylindrical functions and expansion
coefficients involved (for the latter such expansions do not
seem to have been given in the published literature so far,
but we comment on these in Appendix A).

\vspace{3cm}
\noindent
{\large {\bf Acknowledgements}}
\vspace{0.2cm}

D.K.P, S.T. and
J.-z. Z. are indebted to the Deutsche Forschungsgemeinschaft
 (Germany) for financial
support of visits to Kaiserslautern; the work of J.-z.Z.
has also been supported in part by the National Natural Science
Foundation of China under Grant No. 19674014
 and the Shanghai Education Development
Foundation.

\newpage

\begin{appendix}{\centerline{\bf Appendix A}}
\setcounter{equation}{0}
\renewcommand{\theequation}{A.\arabic{equation}}
In ref. \cite{16} on the absorptivity of the $D3$--brane of the
dilaton--axion system it was shown that the $S$--matrix for scattering
of a massless scalar field off the brane is given by
\begin{equation}
S=\frac{(R-\frac{1}{R})e^{-i\pi l}}{Re^{i\nu\pi}-\frac{e^{-i\nu\pi}}{R}}
\label{A.1}
\end{equation}
where
$$
R=\frac{M^{(1)}_{-\nu}(0,h)}{M^{(1)}_{\nu}(0,h)},
$$
$M^{(1)}_{\nu}(z,h)$ being the modified Mathieu function expanded in terms of
Bessel functions, i.e.
$$
Me_{\nu}(0,h)M^{(1)}_{\nu}(z,h)=\sum^{\infty}_{r=-\infty}c^{\nu}_{2r}(h^2)
J_{\nu+2r}(2h\cosh z)
$$
(an expansion with better convergence to use in practice is
 one in terms of products of Bessel functions as shown in ref.\cite{16})
where $Me_{\nu}(z,h)$ is the Fourier or Floquet solution of
the Mathieu equation. In
the published literature the coefficients $c^{\nu}_{2r}(h^2)$
have only been considered as power series in rising powers of $h^2$,
and consequently were used in ref. \cite{16} in the small $h^2$ or low energy
domain.  It would be very interesting to make
the transition to the large--$h^2$ or high energy case
 directly from this expression by developing
large--$h^2$ asymptotic expansions of the
Mathieu function Fourier coefficients $c^{\nu}_{2r}(h^2)$
(for the Bessel functions the corresponding expansions are
known).  We know of no
publication where such expansions have been given,
but one of us (M.--K.) remembers from private communication
with the author of ref.\cite{dingle}
that these Stokes--type asymptotic expansions
can indeed be
obtained. One writes the recurrence relation of the coefficients
(cf. \cite{25}, p.106)
\begin{equation}
c_{2\rho +2}+c_{2\rho -2} =\frac{[\lambda-(\nu+2\rho)^2]}{h^2}c_{2\rho}
\label{A.2} 
\end{equation}
(the Mathieu equation being $y^{\prime\prime}+(\lambda-2h^2\cos 2x)y=0$).
For $|h^2|\rightarrow\infty$ this implies
$$
c_{2\rho +2}\propto i^{(2\rho+2)/2}
$$
Setting
$$
c_{2\rho +2}\equiv b_{\rho+1},\;\;\;\; b_{\rho}=i^{\rho}\beta_{\rho}
$$
we have
\begin{eqnarray}
b_{\rho +1}+b_{\rho -1}&=&\frac{[\lambda-(\nu+\rho)^2]}{h^2}b_{\rho},\nonumber\\
{\beta}_{\rho +1}+{\beta}_{\rho -1}
&=&\frac{-i[\lambda-(\nu+\rho)^2]}{h^2}{\beta}_{\rho}
\label{A.3}
\end{eqnarray}
>From this we deduce that the next approximation
to $c_{2\rho +2}$ is obtained from 
\begin{equation}
{\beta}_r= 1 +\frac{i}{h^2}\sum^r_{\rho =0}\bigg[(\rho +\nu)^2-\lambda\bigg]
\label{A.4}
\end{equation}
The sums on the right hand side can be evaluated. E.g.
$$
\sum^r_{\rho=0}{\rho}^2=1^2+2^2+3^2+\cdot\cdot\cdot +r^2=\frac{1}{6}r(r+1)(2r+1)
$$
so that one obtains
\begin{equation}
{\beta}_r=1+\frac{i}{h^2}\bigg[\frac{r(r+1)(2r+1)}{6}
+2\nu\frac{r(r-1)}{2}+{\nu}^2-\lambda\bigg]
\label{A.5}
\end{equation}
Proceeding in this way one can indeed obtain the desired asymptotic
expansion of the coefficients $c_{2\rho}$.
(In fact the asymptotic expansion of the Bessel function --
similar to that of a linear combination of Hankel functions --
can be obtained from its recurrence relation in a very
similar way).

\end{appendix}

\vspace{1cm}

\begin{appendix}{\centerline{\bf Appendix B}}
\setcounter{equation}{0}
\renewcommand{\theequation}{B.\arabic{equation}}

For the determination of the large--h behaviour of the Floquet
exponent $\nu$ we make use of results of ref.\cite{28}.
A fundamental pair $y_I, y_{II}$ of
respectively even and odd solutions of the original periodic
Mathieu equation with eigenvalue $\lambda$ defined by
$$
y_I(-z)=y_I(z),\;\; y_{II}(z)=-y_{II}(-z)
$$
can be chosen to satisfy the following boundary conditions (cf. e.g.
\cite{25}, pp.99,100)
$$
y_I(0)=1, \;\; y_{II}(0)= 0, \;\;{y_I}^{\prime}(0)=0,
\;\; {y_{II}}^{\prime}(0)=1
$$
>From its original defining property the Floquet exponent $\nu$ can then
be shown to be given by (cf. \cite{25}, p.101)
\begin{equation}
\cos\pi\nu=y_I(\pi;\lambda, h^2)
\label{B.1}
\end{equation}
so that (cf.\cite{25}, p. 100)
\begin{equation}
\cos\pi\nu+1=2y_I(\pi/2;\lambda, h^2){y_{II}}^{\prime}(\pi/2;\lambda, h^2)
\label{B.2}
\end{equation}
The solutions $y_I(z),  y_{II}(z)$ can be identified with the
large--h solutions $ce, se$ of ref.\cite{28}(there eqs.(64))in terms
of functions $A(z), {\bar A}(z)$ as in eq.(\ref{84}) above with normalization
constants $N_0, N_0^{\prime}$, i.e. in leading order
$$
ce(o)=2N_0A(0),\;\; se^{\prime}(0)=4h{N_0}^{\prime}A(0)
$$
from which we deduce in leading order for large $|h|$ that
$$
N_0=2^{-3/2},\;\; {N_0}^{\prime}=2^{-5/2}/h
$$
Eqs.(65) of ref.\cite{28} give the large--h expansions of 
$y_I(\pi/2;\lambda, h^2)$ and ${y_{II}}^{\prime}(\pi/2;\lambda, h^2)$.
Inserting these multiplied by the appropriate normalization
constants into eq.(\ref{B.2}) and retaining the
dominant terms for large $|h|$
we obtain
\begin{equation}
\cos\pi\nu +1 = \frac{\pi e^{4h}}{(8h)^{q/2} }
\bigg[\frac{1+\frac{3(q^2+1)}{64h}}
{\Gamma[\frac{3}{4}-\frac{q}{4}]\Gamma[\frac{1}{4}-\frac{q}{4}]}
+O(\frac{1}{h^2})\bigg]
\label{B.3}
\end{equation}
in agreement with a result cited in ref.\cite{25}(p.210) from \cite{30}
with logarithmic corrections.  We, however, see no such
logarithmic terms in the simpler formulation of ref.\cite{28}.
The relation (\ref{B.3}) we rediscovered here has 
practically been unknown, largely in view of the
difficulty to extract it
from the complicated considerations of ref.\cite{30}. Our derivation
above is simple and closes a difficult gap which
the author of ref.\cite{26} commented upon
with the
words: ``It is not likely at this stage that an analytic relation
will ever be found connecting (our) $\nu$ and $\gamma$
to (our) $a^2$ and $h^2$''. Our search of later literature
did not uncover other derivations. The main source summarizing
more recent developments in the field of the Mathieu equation is
ref.\cite{meix}.

\end{appendix}

\newpage

\centerline{\bf Figure Captions}

\vspace{0.9cm}

\noindent
{\bf Figure 1}

The function $q(h)$ plotted versus $h$, which, of course,
is valid only away from $h=0$.  The plot should be compared
with graphs in ref.\cite{26} where a similar but less convenient
quantity is used.

\vspace{0.5cm}
\noindent
{\bf Figure 2}

The absorptivity $A(l,h)$ for $l=0$.

\vspace{0.5cm}
\noindent
{\bf Figure 3}

The absorptivity $A(l,h)$ for $l=1$.

\vspace{0.5cm}
\noindent
{\bf Figure 4}

The absorptivity $A(l,h)$ for $l=2$.

\vspace{0.5cm}
\noindent

\newpage
\epsfysize=10cm \epsfbox{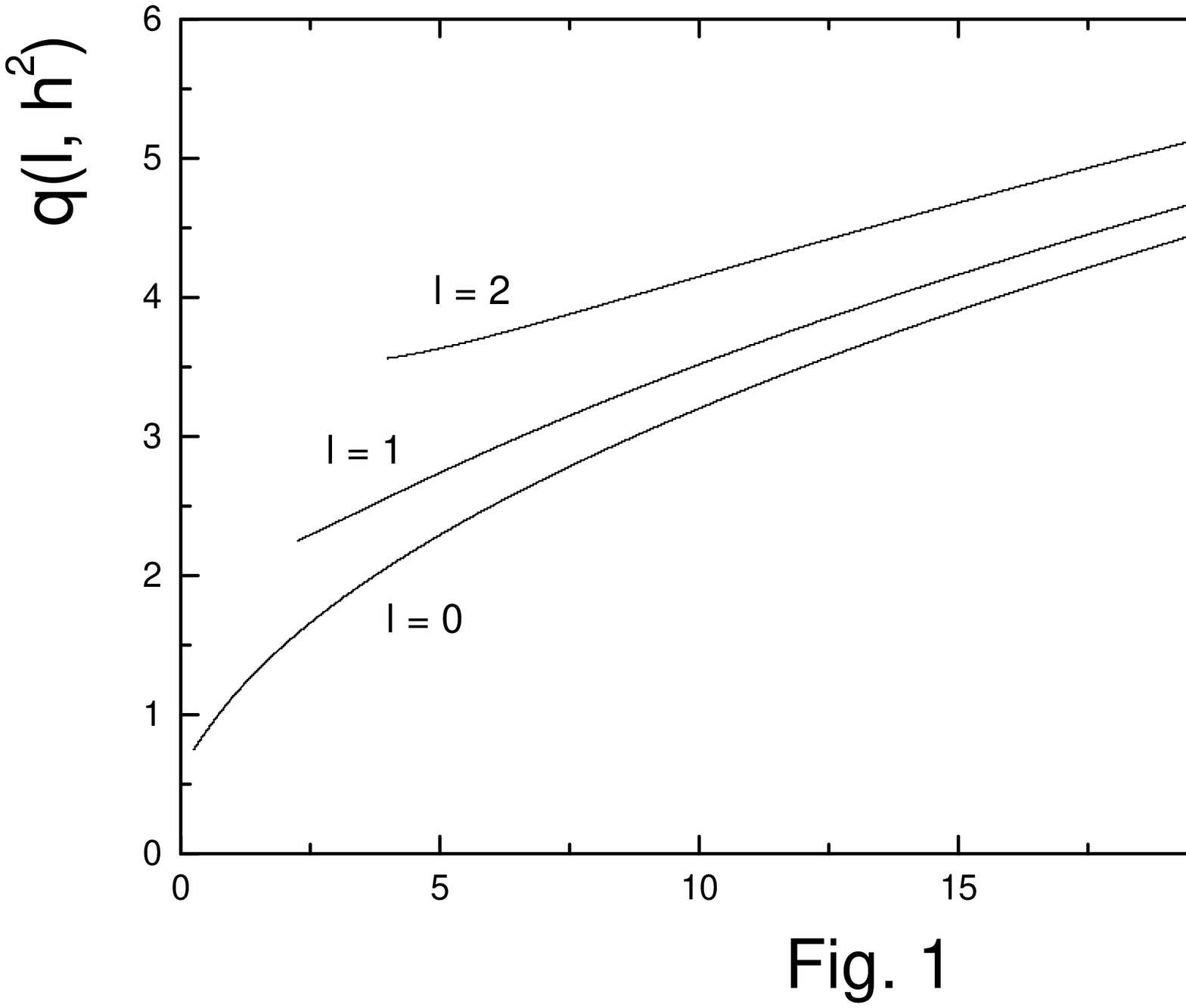}
\newpage
\epsfysize=10cm \epsfbox{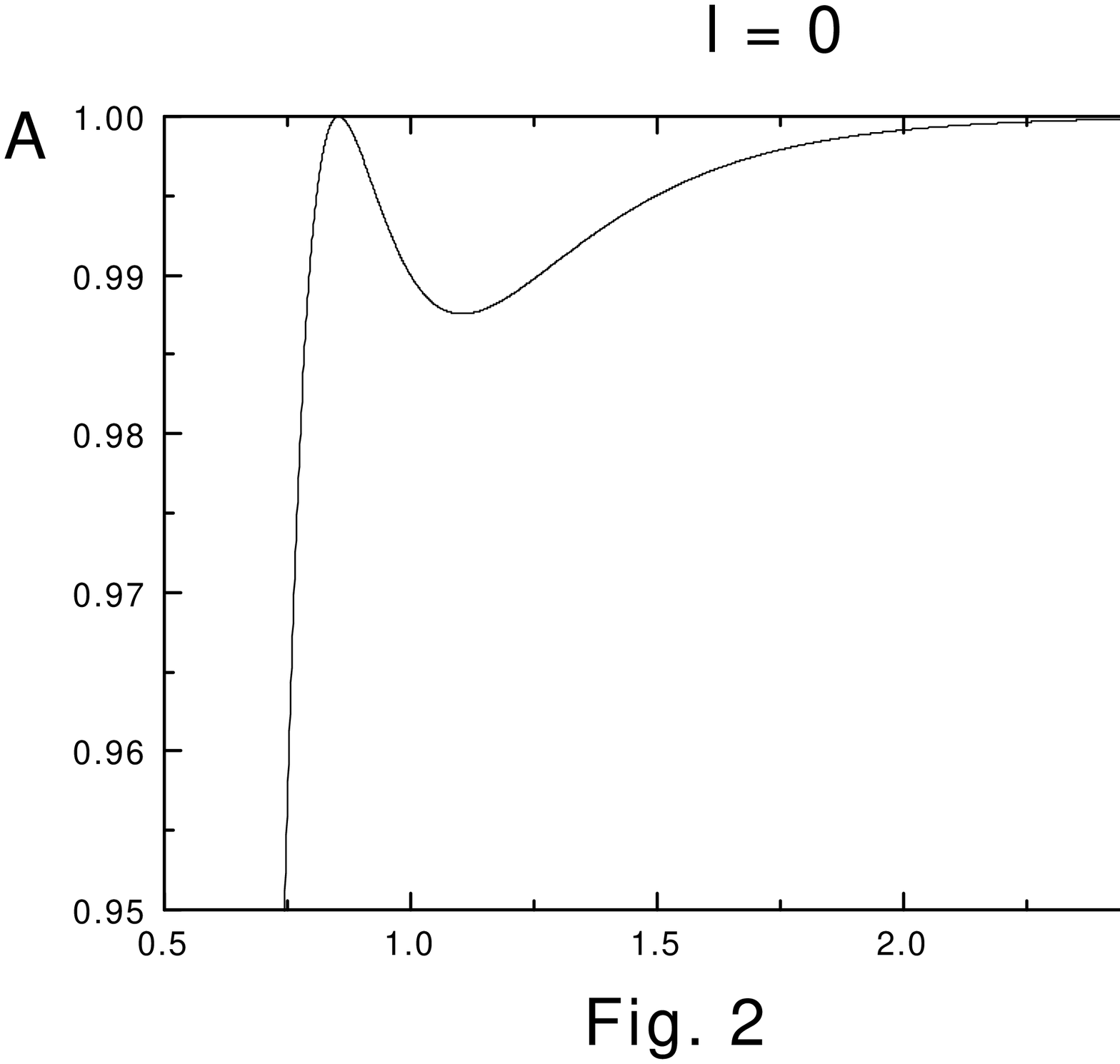}
\newpage
\epsfysize=10cm \epsfbox{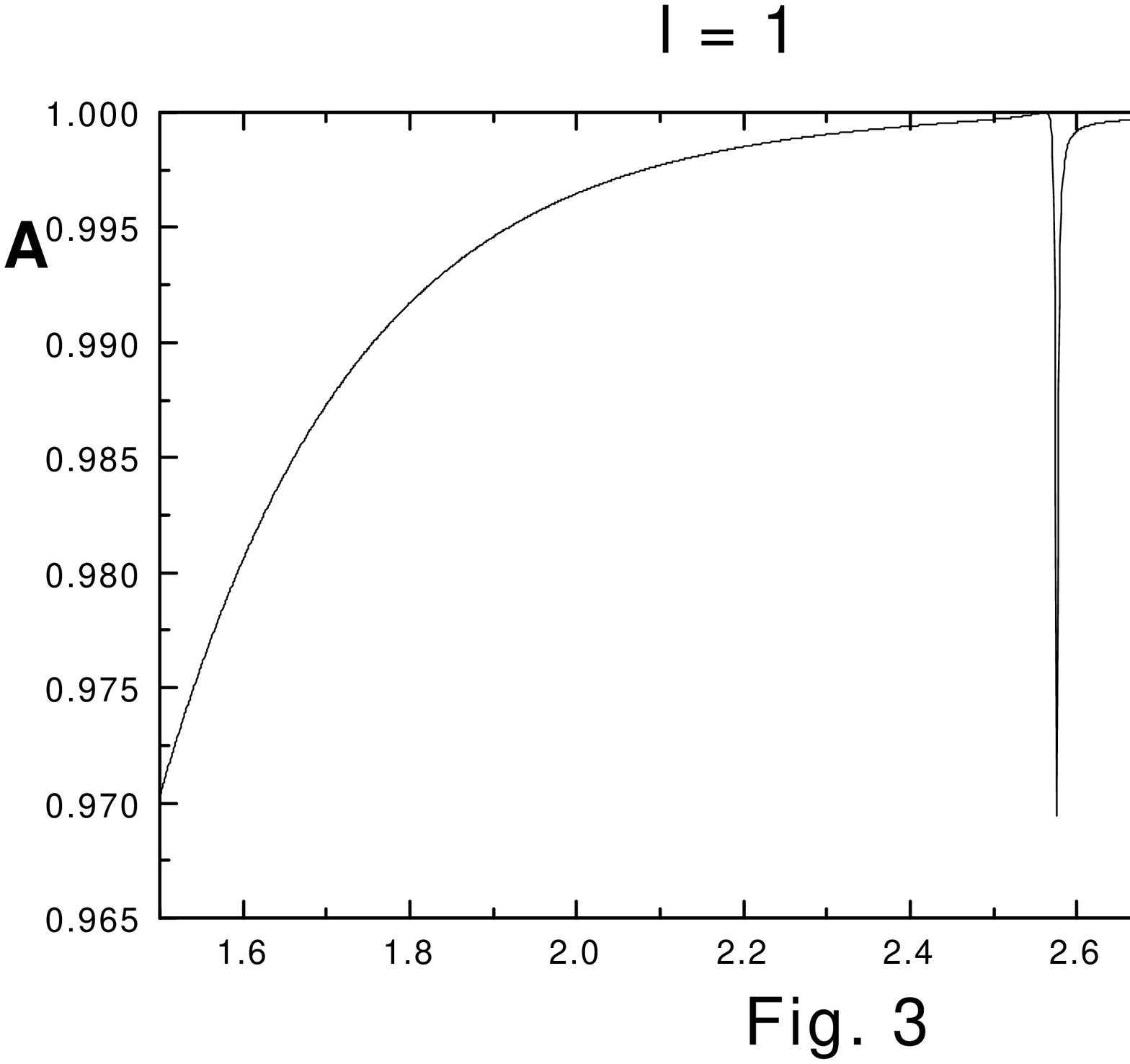}
\newpage
\epsfysize=10cm \epsfbox{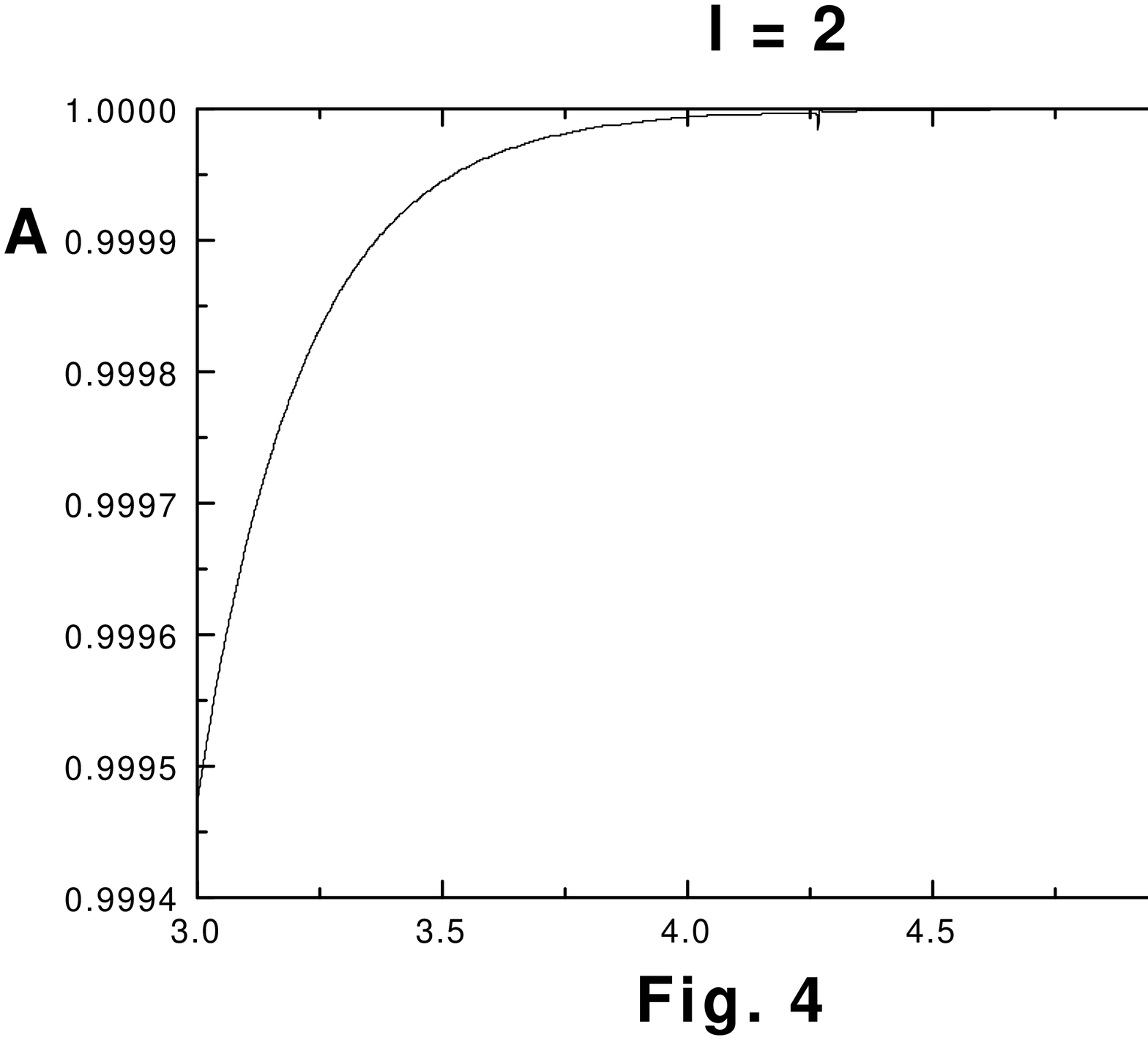}


\vspace{5cm}

\begin{thebibliography}{99}
\bibitem{1} G.W. Gibbons, {\it Lecture at CECS},
Proc. ed. by C. Teitelboim and J. Zanelli, Santiago, Chile,
 August 1997, hep--th/9801106.
\bibitem{2} C.G. Callan and J.M. Maldacena, {\it Brane
 dynamics from the Born--Infeld action}, Nucl. Phys. {\bf B513} (1998) 198,
hep--th/9708147.
\bibitem{3} J. Polchinski, {\it Tasi lectures on $D$-branes}, hep--th/9611050.
\bibitem{4} G.W. Gibbons,
{\it Born--Infeld particles and Dirichlet p--branes}, Nucl. Phys.
 {\bf B514} (1998) 603, hep--th/9709027.
\bibitem{5} A. Hashimoto, {\it The shape of branes pulled by strings},
Phys. Rev. {\bf D57} (1998) 6441,  hep--th/9711097.
\bibitem{6} S. Lee, A. Peet and L. Thorlacius, {\it Brane waves and strings},
Nucl. Phys. {\bf B514} (1998) 161, hep--th/9710097.
\bibitem{7} D. Brecher, {\it BPS States of the
 non--abelian Born--Infeld action}, Phys. Lett. {\bf B442} (1998) 117,
 hep--th/
9804180.
\bibitem{8} S.B. Giddings,{\it Scattering ripples from branes},
Phys. Rev. {\bf D55} (1997) 6367, hep--th/9612022.
\bibitem{9} H.J.W. M\"uller--Kirsten, 
Jian--zu Zhang and D.H. Tchrakian, Int. J. Mod.
Phys. {\bf A5} (1990) 1319. See also
 P. Goddard and D.I. Olive, Rep. Progr. Physics
{\bf 41} (1978) 1357.
\bibitem{10} D. Chruscinski,{\it  Point charge in Born--Infeld
 electrodynamics}, Phys. Lett. {\bf A240} (1998) 8,  hep--th/9712161.
\bibitem{11} A. A. Chernitskii, {\it Nonlinear
 electrodynamics with singularities (Modernized
Born--Infeld electrodynamics)},
Helv. Phys. Acta {\bf 71} (1998) 274,
 hep--th/9705075 v8.
\bibitem{12} S. Gonorazky, C. Nunez, F.A. Schaposnik and 
G. Silva,{\it 
 Bogomol'nyi
bounds and the supersymmetric Born--Infeld theory},
Nucl. Phys. {\bf B531} (1998) 168,
 hep--th/9805054.
\bibitem{13} S.S. Gubser and A. Hashimoto, {\it Exact absorption
probabilities for the $D3$--brane},
Comm. Math. Phys. {\bf 208} (1999) 325, hep--th/9805140.
\bibitem{14} M. Cvetic, H. L\"u, C.N. Pope and T.A. Tran, {\it
Exact absorption probability in the extremal six--dimensional
dyonic string background},
Phys. Rev. {\bf D59} (1999) 126002, hep--th/9901002.
\bibitem{15}M. Cvetic, H. L\"u and J.F. Vazquez--Poritz, {\it Absorption
by extremal $D3$--branes}, hep--th/0002128;
 M. Cvetic, H. L\"u and J. F. Vazquez--Poritz, {\it Massive--scalar
absorption by extremal $p$--branes},
Phys. Lett. {\bf B462} (1999) 62, hep--th/9904135.
\bibitem{kleb97} I.R. Klebanov, {\it World--volume
approach to absorption by non--dilatonic branes},
Nucl. Phys. {\bf B496} (1997) 231, hep--th/9702076v2.
\bibitem{kleb98} S.S. Gubser, I.R. Klebanov and
A. A. Tseytlin, {\it Coupling constant dependence in the
thermodynamics of $N=4$ supersymmetric Yang--Mills theory},
Nucl. Phys. {\bf B534} (1998) 202,  hep--th/9805156.
\bibitem{kleb99} S.S. Gubser, I. R. Klebanov, M. Krasnitz
and A. Hashimoto, {\it Scalar absorption and the breaking
of world volume conformal invariance}, Nucl. Phys.
{\bf B526} (1998) 393, hep--th/9803023.
\bibitem{16} R. Manvelyan, H.J.W. M\"uller--Kirsten, J.--Q. Liang
and Yunbo Zhang,
{\it Absorption cross section of scalar field in
supergravity background}, Nucl. Phys. {\bf B579} (2000) 177,
hep--th/0001179.
\bibitem{singpot} Since metrics of other supergravity
theories lead to singular potentials with 
powers different from that of the case considered
here (i.e. $1/r^4$), we cite some references
which deal with the general case: 
G. Tiktopoulos and S. B. Treiman, {\it Weak--coupling limit
for scattering by strongly singular potentials}, Phys. Rev. {\bf 134} 
(1964) 844;L. Bertocchi, S. Fubini and G. Furlan, {\it The
theory of scattering by singular potentials}, Nuovo Cimento
{\bf 35} (1965) 633; R.A. Handelsman, Y.--P. Pao and J.S. Lew,
{\it Low--energy scattering by long--range singular potentials},
Nuovo Cimento {\bf LV A} (1968) 453;
A. Paliov and S. Rosendorf,
{\it High--energy phase shifts produced by repulsive singular
potentials}, J. Math. Phys. {\bf 8} (1967) 1829.
\bibitem{kleb} I.R. Klebanov,
{\it World volume approach to absorption by non--dilatonic
branes}, Nucl. Phys.
 {\bf B496} (1997) 231, hep--th/9702076,;
 I.R. Klebanov, W. Taylor IV and M. Van Raamsdonk,
{\it Absorption of dilaton partial waves by $D3$ branes}, Nucl. Phys.
{\bf B560} (1999) 207,
hep--th/9905174.
\bibitem{park} D.K. Park and H.J.W. M\"uller--Kirsten, {\it Universality
or non--universality of absorption cross sections for extended
objects}, Phys. Lett. {\bf B} (2000), to be published, hep--th/0008215.
\bibitem{17} P. F. Byrd and M. D. Friedman, {\it Handbook
 of elliptic integrals for
engineers and scientists}, (Springer, 1971), second ed., formulae 110.06 and 260.75.
\bibitem{18} D. Bak, J. Lee and H. Min, {\it Dynamics
 of BPS states in the Dirac--Born--Infeld theory}, Phys. Rev. {\bf D59} (1999) 045011,
hep--th/9806149.
\bibitem{19}K. G. Savvidy and G.K. Savvidy, {\it Von Neumann boundary
conditions from Born--Infeld dynamics}, Nucl.Phys. {\bf B561} (1999)
117, hep--th/9902023; G.K. Savvidy,{\it Electromagnetic dipole
radiation of oscillating $D$ branes},
Nucl. Phys. Proc. Suppl. {\bf 88} (2000) 152, hep--th/0001098;
C. G. Callan and A. G\"uijosa, {\it Undulating strings
and gauge theory waves}, Nucl. Phys. {\bf B565} (2000) 157,
hep--th/9906153; C.G. Callan, A. G\"uijosa, K.G.
Savvidy and O. Tafjord, {\it Baryons and flux tubes in
confining gauge theories from brane actions}, hep--th/9902197.
\bibitem{20} H. H. Aly, H.J.W. M\"uller--Kirsten and N. Vahedi--Faridi,
{\it Scattering by singular potentials with a perturbation--theoretical
introduction to Mathieu functions},
 J. Math. Phys.{\bf 16}
(1975) 961--970.
\bibitem{21} R. M. Spector,{\it Exact
solution of the Schr\"odinger equation for
inverse fourth power potential}, J.Math. Phys.{\bf 5}(1964) 1185.
\bibitem{22} N. Limic, Nuovo Cimento {\bf 26} (1962) 581.
\bibitem{23} H. H. Aly, W. G\"uttinger and H. J. W. M\"uller--Kirsten,
{\it Singular interactions}, 
Lectures on Particles and Fields, ed. H. H. Aly, Gordon and Breach (1970), p. 247.
\bibitem{24} G. Esposito,{\it  Scattering from singular
 potentials in quantum mechanics}, J. Phys. {\bf A31} (1998) 9493,
hep--th/9807018.
\bibitem{25} J. Meixner and F. W. Sch\"afke, {\it Mathieusche
 Funktionen und Sph\"aroidfunktionen},
(Springer, 1953).
\bibitem{26} G. H. Wannier,
{\it Connection formulas between the solutions of Mathieu's
equation}, Quart. Appl. Math. {\bf 11}(1953) 33.
\bibitem{27} H.H.Aly, H.J.W. M\"uller and N. Vahedi--Faridi,
{\it Some remarks on scattering by singular potentials}, Lett. Nuovo
Cimento {\bf 2} (1969) 485.
\bibitem{28}R. B. Dingle and H.J.W. M\"uller,
{\it Asymptotic expansions of Mathieu functions
and their characteristic numbers}, J. reine angew. Math.
{\bf 211} (1962) 11.
\bibitem{29} A.A. Tseytlin, {\it Self--duality of Born--Infeld action and
Dirichlet 3--brane of type IIB superstring theory}, Nucl. Phys. {\bf B469}
(1996) 51, hep--th/9602064; M. Aganagic, J. P. Park,
C. Popescu and J. H. Schwarz,{\it Dual $D$--brane actions}, hep--th/9702133.
\bibitem{30}R.E. Langer,{\it The solutions of the Mathieu equation
with a complex variable and at least
one parameter large}, Trans. Amer. Math. Soc. {\bf 36} (1934)
637 -- 695.
\bibitem{dingle} R. B. Dingle, {\it Asymptotic expansions: Their derivation and interpretation}, (Academic, 1973).
\bibitem{meix} J. Meixner, F.W. Sch\"afke and G. Wolf, {\it Mathieu
functions and spheroidal functions and their mathematical foundations:
Further studies}, (Springer, 1980).

\end{thebibliography}
\end{document}